\documentclass[11pt]{article}
\pdfoutput=1
\usepackage{jcapmod}
\usepackage{shorthand}

\usepackage{mathtools}
\usepackage{booktabs}
\usepackage[english]{babel}
\usepackage{amsmath,amssymb,amsbsy,amstext, amsthm, simplewick, amsfonts}
\usepackage{hyperref}
\usepackage{graphicx}
\usepackage[small]{caption}
\usepackage{siunitx}
\usepackage{upgreek}
\usepackage{framed}
\usepackage{wrapfig}
\usepackage{multirow}
\usepackage{bbm}
\usepackage[svgnames,dvipsnames,x11names]{xcolor}
\usepackage[utf8x]{inputenc}
\usepackage{selinput}
\usepackage{bm}
\usepackage{float}
\usepackage{geometry}
\usepackage{yfonts}
\usepackage{caption}
\usepackage{subcaption}
\usepackage{sidecap}
\usepackage{longtable}
\usepackage{anyfontsize}
\usepackage{dsfont}

\usepackage[framemethod=default]{mdframed}
\newmdenv[skipabove=7pt,
skipbelow=7pt,
rightline=false,
leftline=false,
topline=false,
bottomline=false,
backgroundcolor=gray!10,
linecolor=gray,
innerleftmargin=5pt,
innerrightmargin=5pt,
innertopmargin=5pt,
innerbottommargin=5pt,
leftmargin=0cm,
rightmargin=0cm,
linewidth=4pt]{eBox}
\newmdenv[skipabove=7pt,
skipbelow=7pt,
rightline=true,
leftline=true,
topline=true,
bottomline=true,
backgroundcolor=white,
linecolor=gray,
innerleftmargin=5pt,
innerrightmargin=5pt,
innertopmargin=5pt,
innerbottommargin=5pt,
leftmargin=0cm,
rightmargin=0cm,
linewidth=1pt]{eBox2}

\definecolor{blue3}{RGB}{31, 119, 180}
\definecolor{red3}{RGB}{	214, 39, 40}
\definecolor{orange3}{RGB}{255, 127, 14}
\definecolor{green3}{RGB}{44, 160, 44}
\definecolor{repBlue}{RGB}{31, 119, 180}
\definecolor{repRed}{RGB}{	214, 39, 40}
\definecolor{repGreen}{RGB}{44, 160, 44}

\renewcommand{\(}{\left(}
\renewcommand{\)}{\right)}
\renewcommand{\[}{\left[}
\renewcommand{\]}{\right]}

\def\be{\begin{equation}}
\def\ee{\end{equation}}
\newcommand{\bea}{\begin{eqnarray}}
\newcommand{\eea}{\end{eqnarray}}
\def\fnl{f_{\rm NL}}
\def\vp{\varphi}

\usepackage{colortbl}
\definecolor{lightgreen}{cmyk}{0.2, 0, 0.2, 0.2}
\definecolor{lightgray}{cmyk}{0.1,0.2,0,0.1}
\definecolor{lightgray2}{cmyk}{0.1,0.1,0,0.1}

\makeatletter
\newlength{\apb@width}
\newcommand{\autoparbox}[2][c]{\settowidth{\apb@width}{#2}\parbox[#1]{\apb@width}{#2}}

\makeatother


\def\beq{\begin{equation}}
\def\eeq{\end{equation}}

\newcommand{\fNL}{{f_\mathrm{NL}}}
\newcommand{\ells}{{\ell_1 \ell_2 \ell_3}}
\newcommand{\sigmasnr}{{\sigma_\mathrm{SNR}}}

\allowdisplaybreaks[1]
\begin{document}

\newgeometry{top=2cm, bottom=2cm, left=2.9cm, right=2.9cm}

\begin{titlepage}
\setcounter{page}{1} \baselineskip=15.5pt 
\thispagestyle{empty}

\begin{center}
{\fontsize{14.9}{18} \bf Searching for  Cosmological Collider in  the Planck CMB Data}\\
\end{center}

\vskip 20pt

\begin{center}
\noindent
{\fontsize{12}{18}\selectfont Wuhyun Sohn$^1$, Dong-Gang Wang$^2$, James R. Fergusson$^2$, and  E. P. S. Shellard$^2$}
\end{center}

\vskip 20pt

\begin{center}
  \vskip8pt
   {$^1$ \fontsize{10.7}{18}\it Korea Astronomy and Space Science Institute, Daejeon 34055, South Korea
}

  \vskip8pt
{$^2$\fontsize{10.7}{18}\it Centre for Theoretical Cosmology, Department of Applied Mathematics and Theoretical Physics\\ University of Cambridge,
Wilberforce Road, Cambridge, CB3 0WA, UK}

\end{center}

\vskip 20pt

%
%

\vspace{0.4cm}
 \begin{center}{\bf Abstract} 
 \end{center}
 \noindent
In this paper, we present the first comprehensive CMB data analysis of cosmological collider physics. 
New heavy particles during inflation can leave imprints in the primordial correlators which are observable in today's cosmological surveys. 
This remarkable detection channel provides an unsurpassed opportunity to probe new physics at extremely high energies.
Here we initiate the search for these relic signals in the cosmic microwave background (CMB) data from the Planck legacy release.
On the theory side, guided by recent progress from the cosmological bootstrap, we first propose a family of analytic bispectrum templates that incorporate the distinctive signatures of cosmological collider physics. Our consideration includes the oscillatory signals in the squeezed limit, the angular dependence from spinning fields, and several new shapes from nontrivial sound speed effects. 
On the observational side, we apply the recently developed pipeline, CMB Bispectrum Estimator (CMB-BEST), to efficiently analyze the three-point statistics and search directly for these new templates in the Planck 2018 temperature and polarization data. 
We report stringent CMB constraints on these new templates. Furthermore, we perform parameter scans to search for the best-fit values with maximum significance.
For a benchmark example of collider templates, we find $\fnl=-91\pm40$ at the $68\%$ confidence level. After
accounting for the look-elsewhere effect, the biggest adjusted significance we get is $1.8\s$.
In general, we find no significant evidence of cosmological collider signals in the Planck data.
However, our innovative analysis,  together with the recent work \cite{Cabass:2024wob} using the BOSS data,  sets the stage for probing cosmological collider  and  demonstrates the potential for discovering new heavy particles during inflation in forthcoming cosmological surveys.

\noindent

\end{titlepage}

\newpage

\restoregeometry
\setcounter{tocdepth}{3}
\setcounter{page}{1}
\tableofcontents

\vskip 16pt

 \hrule

\vskip 25pt

\section{Introduction}
\label{sec:intro}

One remarkable aspect of inflationary cosmology is that the earliest stage of our Universe also serves as a natural laboratory for high energy physics, where the effects of the microscopic quantum world during inflation are imprinted on the macroscopic classical spacetime, which can be probed later through astronomical observations \cite{Meerburg:2019qqi,Achucarro:2022qrl}.
One of the most promising channels for detecting signatures of the early universe lies in the higher-order cosmological correlators of primordial fluctuations that are formed during inflation.
As these spatial correlations capture deviations from Gaussian statistics, they are usually known as primordial non-Gaussianity (PNG), and for this reason they have been an important observational target in on-going and upcoming experiments, such as those mapping the cosmic microwave background (CMB) and for large scale structure (LSS) surveys. 
Given that inflation may reach the {\it highest} energy densities that are observationally accessible in nature ($\lesssim 10^{13}$GeV), the search for PNG provides a unique and exciting opportunity to probe fundamental physics at energy scales far beyond the reach of  terrestrial experiments.

\vskip4pt
This confrontation between observational cosmology and high energy theory finds a particularly lucid exposition in the ``Cosmological Collider Physics" program~\cite{Chen:2009zp, Baumann:2011nk, Noumi:2012vr, Arkani-Hamed:2015bza}. The analogy is drawn with particle accelerators where interactions with intermediate particles is imprinted in the scattering amplitudes determined in flat spacetime.  For new heavy particles during inflation, likewise, they can mediate specific correlations of primordial fluctuations, with their mass and spin leading to distinctive signatures in the shape of PNG. Therefore, this provides the opportunity to do particle spectroscopy by measuring the statistics of primordial fluctuations. 
Considering the rich phenomenology available in cosmological observations  and their potentially deep implications for particle physics, theoreticians have extensively studied various possibilities of the cosmological collider physics in the past decade \cite{Chen:2009we, Assassi:2012zq, Chen:2012ge, Pi:2012gf,  Chen:2015lza, Lee:2016vti, Chen:2016uwp, Chen:2016hrz, Chen:2017ryl, Kehagias:2017cym, Kumar:2017ecc, An:2017hlx, An:2017rwo, Baumann:2017jvh, Chen:2018xck,Kumar:2018jxz, Bordin:2018pca,  Kim:2019wjo, Alexander:2019vtb,  Wang:2019gbi, Wang:2019gok, Aoki:2020zbj, Lu:2021wxu, Wang:2021qez, Tong:2021wai, Cui:2021iie, Tong:2022cdz, Reece:2022soh,  Chen:2022vzh, Qin:2022lva,   Niu:2022quw, Werth:2023pfl, Chen:2023txq, Xianyu:2023ytd, Chakraborty:2023qbp,Jazayeri:2023xcj, Aoki:2020wzu, Pinol:2023oux, McCulloch:2024hiz, Wu:2024wti}. 

\vskip4pt
The recent development of the cosmological bootstrap has further enhanced theoretical efforts along this direction \cite{Arkani-Hamed:2018kmz,Baumann:2019oyu,Baumann:2020dch, Arkani-Hamed:2017fdk, Benincasa:2018ssx, Sleight:2019mgd, Sleight:2019hfp, Goodhew:2020hob, Cespedes:2020xqq, Pajer:2020wxk, Jazayeri:2021fvk, Bonifacio:2021azc, Melville:2021lst, Goodhew:2021oqg, Pimentel:2022fsc, Jazayeri:2022kjy,  Baumann:2022jpr,  Qin:2022fbv, Salcedo:2022aal, Wang:2022eop,Qin:2023ejc,De:2023xue,Stefanyszyn:2023qov,DuasoPueyo:2023viy,Cespedes:2023aal,Arkani-Hamed:2023bsv,Arkani-Hamed:2023kig,Bzowski:2023nef,Chen:2023iix, Donath:2024utn, Fan:2024iek,  melville2024sitter}.
This new approach aims to directly determine the forms of cosmological correlators from a set of fundamental principles, such as symmetry, unitarity and locality.  For this reason, the bootstrap methodology makes it possible to carve out the theory space in a model-independent way, and achieve a systematic classification of the possible categories for all the  predictions.
In addition, the approach also provides powerful computational tools. For the first time, we are able to compute the shape function of cosmological colliders for any kinematics, and identify the full analytic structure of the corresponding correlators.
These latest advances have inspired a more complete survey revealing non-Gaussianity predictions with new features and large signals that can be tested even using currently available data.

\vskip4pt
At the observational frontier the searches for cosmological collider signals in real data are far from exhausted and discovery potential remains. 
The leading target of PNG has been the bispectrum of primordial curvature perturbations, with most of the data analysis to date focused on simpler templates from multi-field and single field inflation models, such as the local and equilateral shapes.  
The latest and tightest constraints on these non-Gaussianities come from the {\it Planck} CMB experiment \cite{Akrami:2019izv} (see also \cite{Planck:2013wtn,Planck:2015zfm}), which included general methods to cover a much wider array of theoretically-motivated templates \cite{Fergusson:2009nv,Fergusson:2010dm,Bucher_2010}, though only few were directly inspired by cosmological colliders (see, e.g.\ \cite{Sefusatti:2012ye}).
Constraints on PNG have been placed using LSS datasets including the BOSS survey through scale-dependent bias \cite{Dalal:2008,Matarrese:2008,Mueller:2022,Cagliari:2023} or through galaxy power spectrum and bispectrum \cite{Cabass:2022wjy,Cabass:2022ymb,DAmico:2022gki}
albeit being very much weaker than those currently available from the CMB.
For the cosmological collider signal, there are already several forecast studies relevant to future galaxy surveys \cite{Meerburg:2016zdz, MoradinezhadDizgah:2017szk, MoradinezhadDizgah:2018ssw}, but no comprehensive CMB data analysis has been performed yet.
One major difficulty is that the shape functions with collider signals are very complicated in general due to oscillations and various types of special functions. This poses a number of technical challenges for the data analysis in order to extract the signal of physical interest.

\vskip4pt
The CMB bispectrum data analysis is, in general, computationally challenging because a naive implementation scales as $O(\ell_\mathrm{max}^5)$, with $\ell_\mathrm{max}\approx 2500$ for \textit{Planck}. Three main approaches have been utilised in the literature to mitigate this problem: the KSW estimator \cite{Komatsu:2003iq,Creminelli2006limits,Yadav2007,Senatore:2009gt,Smith:2006ud}, exploiting the separability of simple approximate templates reduces high-dimensional integrals into products of lower dimensional integrals; the Modal estimator \cite{Fergusson:2009nv,Fergusson:2010dm,Fergusson:2014gea}, which expands arbitrary bispectrum shapes to high precision using separable basis functions; and the related binned bispectrum estimator \cite{Bucher_2010,Bucher:2015ura}, which uses multipole bins to compress the data. However, the KSW and binned estimators are not well suited to an extensive study of cosmological colliders, because the resulting bispectrum shapes tend to be non-separable and highly oscillatory.

\vskip4pt
In this work, we use the high-resolution CMB bispectrum estimator with a publicly available Python code, named CMB-BEST (CMB Bispectrum ESTimator) \cite{Sohn:2023fte}. The formalism combines the advantages of two conventional methods (KSW and Modal) using a flexible set of modal basis functions in the primordial space and it has been extensively validated against the Modal bispectrum pipeline used for the Planck analysis \cite{Fergusson:2010dm,Akrami:2019izv}.  All computationally expensive steps are precomputed at a given modal resolution into a data file provided with the code, so that the users can obtain the \textit{Planck} 2018 constraints on arbitrary shape functions of interest rapidly (within a minute on a laptop). Because of CMB-BEST's flexible mode expansions and its speed and resolution, it is an ideal tool for studying cosmological collider signals.

\begin{table}[b]
    \centering
    \renewcommand{\arraystretch}{1.5} 

    \begin{tabular}{lccccc}
    \toprule
    Shape &    Template &      $\fNL$ (68\% CL) & Raw S/N & Adjusted S/N & Section \\
    \midrule
        Light scalar exchange \cite{Chen:2009zp} & \eqref{qsf} &  $10 \pm 26$ & 0.37 & 0.12 &  \ref{sec:results_QSF_SELM_OC}  \\
        Scalar exchange I   & \eqref{scalarI} &  $11 \pm 13$ &                                  0.86 & 0.67 &  \ref{sec:results_QSF_SELM_OC}  \\
        Scalar exchange II   & \eqref{scalarII} & $-91 \pm 40$ &                                 2.3 & 1.8 & \ref{sec:results_QSF_SELM_OC}  \\
        Heavy-spin exchange  & \eqref{eqn:template_SE} & $-59 \pm 32$ &                                 1.9 & 1.2 &  \ref{sec:results_SE_SO}  \\
        Massive spin-2 exchange & \eqref{spin2m} & $-2.1 \pm 1.1$ &                                 1.9 & 0.90 &  \ref{sec:results_SE_SO}  \\
        Equilateral collider \cite{Pimentel:2022fsc} & \eqref{eqcol} &  $-178 \pm 72$ &                                  2.5 & 0.90 & \ref{sec:results_EC_LSC_MS}  \\
        Low-speed collider \cite{Jazayeri:2023xcj} & \eqref{lowcs} & $-9 \pm 10$ &                                 0.89 & 0.29 & \ref{sec:results_EC_LSC_MS}  \\
        Multi-speed PNG \cite{Wang:2022eop} & \eqref{multics} & $-3.1 \pm 2.3$ &                                 1.3 & 0.61 & \ref{sec:results_EC_LSC_MS}  \\
    \bottomrule
    \end{tabular}
    \caption{Summary of the CMB bispectrum constraints presented in this work. Most templates have free parameters, which were set to their best-fit values for the `constraints' column in this Table; refer to the corresponding section for the full constraints. The adjusted signal-to-noise incorporates the look-elsewhere effect, also detailed in the section. }
    \label{tab:summary_constraints}
\end{table} 

\vskip4pt
On the basis of the theoretical survey presented here, we have deployed the CMB-BEST pipeline to perform the first comprehensive CMB search for cosmological collider signals, yielding the most precise measurements to date of the wide array of relevant bispectrum shapes.
The originality of this work is twofold:
\begin{itemize}
    \item We derive a set of cosmological collider bispectrum templates that are relatively simple and sufficiently accurate across the whole observational domain. With the help of the bootstrap approach, we exploit the systematic classification and complete analytic understanding of the bispectrum shapes to capture the relevant possibilities and to simplify the analytical expressions.
In addition, we also endeavour to collate all the various types of shape ansatzes given in the cosmological collider literature. 
Our consideration of the shape functions incorporates: 1) the oscillatory signals from massive fields; 2) the angular-dependence of profiles due to spinning particles; and 3) three additional new shapes caused by the effect of varying the sound speed.
One particular advantage of these simplified bispectrum templates is that it allows us to perform {\it parameter scans} in the present data analysis.
Their analytical expressions are shown in grey boxes in the text and they can also be directly applied to future observational surveys.
    \item We perform a systematic search for these collider templates in the latest Planck temperature and polarization data. Our main results are summarized in Table \ref{tab:summary_constraints}, which lists the measurements of the collider bispectrum templates presented in the different referenced sections of this work. These summary statistics are given as ``constraints'', but actually represent the highest signal-to-noise measurements obtained, that is, the best-fit parameter choices after scans over the available parameter space for which both the theoretical predictions and observational methods are robust.  

    \end{itemize}

To offer a brief overall summary at the outset, we find no significant evidence for cosmological collider signals given the resolution and sensitivity available with the present Planck CMB data, and within the improved theoretical approximations made for this investigation. As the first full analysis of cosmological collider signals, we expect the methodology we have presented here will be directly applicable to forthcoming observations, notably for Simons Observatory and CMB-S4, but also future large-scale structure surveys, which forecasts suggest may become competitive with the CMB.

The structure of the paper is organized as follows.
In Section \ref{sec:shape}, we first review the classification of the primordial bispectrum shapes from the bootstrap perspective, and then based on the exact analytical solution, we propose a set of simplified templates for various types of cosmological collider signals, for the convenience of the data analysis.
More technical details about how to construct the approximated shape functions are left in Section \ref{app:seed}.
In Section \ref{sec:best}, we introduce basic setup and methodology of the python pipeline CMB-BEST.
In Section \ref{sec:test}, we present the CMB constraints on various shape templates of cosmological colliders using the latest Planck data.
We conclude in Section \ref{sec:concl}.

\section{Shapes of the Cosmological Collider}
\label{sec:shape}

In this section, we briefly review the non-Gaussian shape functions of the primordial scalar bispectra in cosmological collider scenarios. We exploit the full analytic understanding of the three-point functions from the bootstrap analysis, and then propose the approximated shape templates for various types of new physics signatures.

\subsection{PNG: A Bootstrap Recap}
The major target of our analysis is the primordial bispectrum of the curvature perturbation $\zeta=(H/\dot\Phi )\phi$ 
\begin{align}
\langle \zeta_{{\bf k}_1} \zeta_{{\bf k}_2} \zeta_{{\bf k}_3}  \rangle = (2\pi)^3 \delta({\bf k}_1+{\bf k}_2+{\bf k}_3) \frac{18}{5}\fnl \frac{S(k_1,k_2,k_3)}{k_1^2 k_2^2 k_3^2} P_\zeta^2~
,
\end{align}
where $ P_\zeta$ is the power spectrum of $\zeta$ and $\fnl$ represents the size of the non-Gaussian signal. 
The information of new physics during inflation is mainly captured by the dimensionless function $S(k_1,k_2,k_3)$.

In our current understanding of PNG, if we assume the inflation theory is (nearly) scale-invariant and weakly coupled, there are three broad classes of scalar bispectra, which correspond to the Feynman diagrams shown in Figure \ref{fig:png}. One remarkable advantage of the bootstrap approach is that, without referencing to specific models, we are allowed to systematically classify all the possible bispectrum shapes based on the minimal assumptions.
The PNG predictions from the three leading scenarios of inflation are summarized as follows.

\begin{itemize}
    \item {\bf Single field inflation}: In this scenario, the scalar bispectrum is generated by the self-interaction of the inflaton. As  the inflaton field is normally expected to be protected by a shift symmetry, we have derivative interactions.  The bootstrap analysis of single field scenario is presented by the {\it boostless bootstrap} \cite{Pajer:2020wxk, Jazayeri:2021fvk, Bonifacio:2021azc}, which starts from an ansatz of rational polynomials and then keeps imposing constraints from scale invariance, Bose symmetry, flat-space limit and locality. In the end the shape functions from derivative interactions can be written into the following form
    \be
    S^{\rm eq}(k_1,k_2,k_3) = \frac{{\rm Poly}_{p+3}(k_T, e_2, e_3)}{k_1k_2k_3 k_T^p}~,
    \ee
    where $k_T=k_1+k_2+k_3$, $e_2=k_1k_2+k_2k_3+k_1k_3$ and $e_3 = k_1 k_2 k_3$ are symmetric polynomials. The index $p$ is an integer related to the number of derivatives in the cubic vertex, while the form of the $(p+3)$-order polynomial in the numerator is fixed by the flat-space amplitude and the manifestly local test.
    From the perspective of effective field theories, the lowest order interactions are given by $\dot\phi^3$ and $\dot\phi(\partial_i\phi)^2$, which may be seen as consequences of integrating out heavy physics with masses much larger than Hubble. As in this case the  contact field interactions are local, and enhanced around (sound) horizon-exit, the resulting bispectra are usually peaked at the equilateral configuration with $k_1=k_2=k_3$, which are known as the {\it equilateral-type non-Gaussianity}.
    
\begin{figure}
    \centering
    \includegraphics[width=0.8\textwidth]{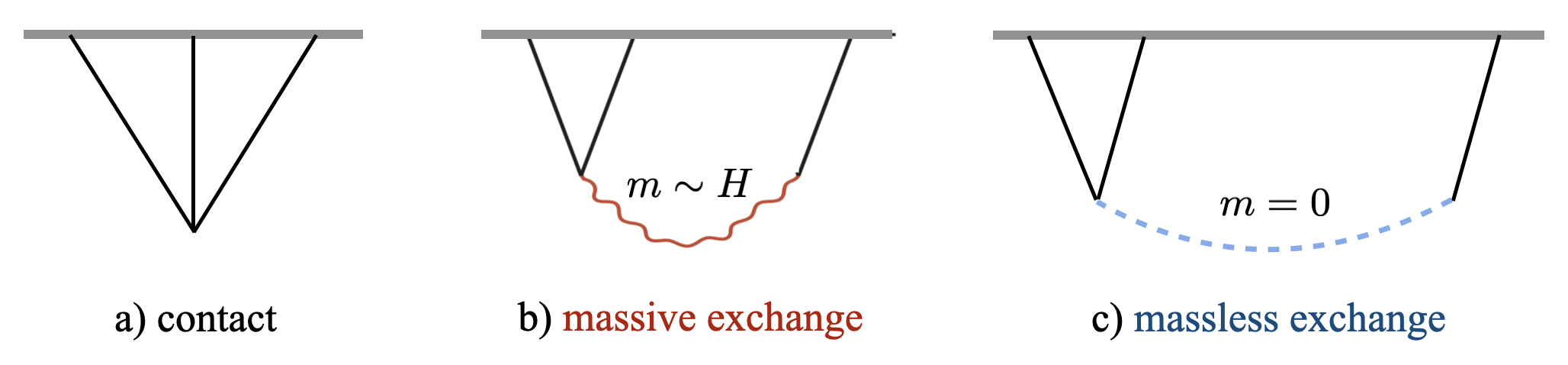}
    \caption{Feynman diagrams for three leading predictions of primordial bispectra from  inflation. In these diagrams, time flows from bottom to top, and the horizontal grey lines correspond to the end of inflation. All the external legs (black lines) are the inflaton fluctuations $\phi$ which are associated to the curvature perturbation $\zeta$. Each diagram describes one scenario of cosmic inflation with distinctive non-Gaussian signatures: a) the contact diagram for single field inflation; b) the massive exchange diagram for cosmological colliders; c) the massless exchange diagram for  multi-field inflation. This work mainly focuses on testing non-Gaussian signals from the second diagram.}
    \label{fig:png}
\end{figure}

    \item {\bf Multi-field inflation}: Non-Gaussianities from additional light scalars during inflation have been extensively studied in literature. The traditional approach is to take the separate universe approximation, and the $\delta N$ formalism there intuitively captures the conversion process, which leads to the famous local shape $S^{\rm loc}=(k_1^3+k_2^3+k_3^3)/(3k_1k_2k_3)$. This type of signal can be understood from the massless exchange diagram shown in the last diagram in Figure \ref{fig:png}, where the intermediate light isocurvature field can travel a long distance before converting into the curvature perturbation. This leads to shape functions peaked in the squeezed limit $k_3\ll k_1\simeq k_2$, which are dramatically different from the single field ones. A recent bootstrap analysis shows that the local-type PNG are consequences of IR divergences in de Sitter space, which in field-theoretic computations lead to mild logarithmic deviation of the shape functions from the standard local ansatz \cite{Wang:2022eop}. 
    \item {\bf Cosmological collider}: roughly speaking, this scenario can be seen as an interpolation between the previous two. As we have argued, when the extra fields are heavy $m\gg H$, they quickly decay into the inflaton which leads to contact interactions in the effective single field inflation. The opposite situation happens for additional light scalars with $m\ll H$  where the long-lived internal state leads to multi-field non-Gaussianities. Their differences can be demonstrated more clearly by taking the squeezed limit of the primordial bispectra:
    \bea \label{eq.scaling}
    \lim_{k_l\ll k_s} S^{\rm eq} & \propto & \frac{k_l}{k_s} \\
     \lim_{k_l\ll k_s} S^{\rm loc} & \propto & \frac{k_s}{k_l} 
    \eea
    In addition to the above single field and multi-field non-Gaussianities, the cosmological collider captures the intermediate regime of massive states $m\sim H$, where the squeezed bispectrum interpolates the two scalings above.
    Furthermore, rich phenomenologies are expected, from which we may be able to extract  information about the mass and spin of the unknown heavy particles during inflation. In the rest of this section, we shall elaborate on this by looking into various possibilities of the shape functions of cosmological colliders.
\end{itemize}

On the observation frontier, there have been many efforts for testing the local template and the equilateral-type shape functions (normally for equilateral and orthogonal templates), using both the CMB data and the LSS surveys.
But few data analysis has been performed for cosmological colliders yet, even though the phenomenology there is richer. 
One major challenge arises from  the theory side, as the analytical forms of predicted shape functions (if exist) are very complicated.
In the rest of this section, we shall  tackle this problem by proposing well-motivated simple templates.

\subsection{Shapes from massive scalar exchange}
\label{sec:scalar}

One characteristic signal of cosmological colliders is the oscillations in the squeezed bispectrum. 
This corresponds to 
the generalized scaling behaviour in the soft limit, which can already been identified in massive scalar exchanges.
Through the Feynman diagram in Figure \ref{fig:png}b), the super-horizon decay of the massive particle during inflation leaves the imprints in  cosmological correlators.
In general, there are two classes depending on the mass of the intermediate particle.
\begin{itemize}
    \item 
When the massive scalar is in the complimentary series ($0<m<3H/2$), the field demonstrates a power-law decay with conformal time after horizon-exit. 
The signatures of massive scalars were first extensively studied in the quasi-single field (QSF) inflation\footnote{In general, the quasi-single field models correspond to the scenario with $m\sim H$ massive scalars, which also includes the $m>3H/2$ case. We use the acronym ``scalar-L" for the  template of the light-mass $m<3H/2$ regime.}, whose squeezed bispectrum takes the following form \cite{Chen:2009zp, Baumann:2011nk, Assassi:2012zq}
\be \label{qsf-sq}
\lim_{k_l\ll k_s} S^{\rm scalar-L}  \propto  \(\frac{k_l}{k_s} \)^{\frac{1}{2}-\nu} ~~~~~~~~~~~~~~
{\rm with}~ \nu=\sqrt{\frac{9}{4}-\frac{m^2}{H^2}} ~~~~~
\ee
with the index $\nu\in (0,3/2)$ given by the mass. For this type of models, an approximated shape template has been proposed in \cite{Chen:2009zp}
\begin{eBox}
\be \label{qsf}
 S^{\rm scalar-L}(k_1,k_2,k_3) =  
 3\sqrt{3\kappa} \frac{N_\nu(8\kappa)}{N_{\nu}({8}/{27})}~, ~~~~ {\rm with}~\kappa=\frac{k_1k_2k_3}{k_T^3}
\ee     
\end{eBox}
where $N_\nu$ is the Neumann function.
In the squeezed limit this template can reproduce  the correct scaling behaviour in \eqref{qsf-sq}. Meanwhile for the non-squeezed kinematics configuration, by varying the mass parameter, the full shape provides an interpolation between two types of templates \cite{Chen:2009zp}: when $\nu\rightarrow 3/2$, it goes back to the local shape from multi-field inflation; when $\nu\rightarrow 0$ we find it approaches the constant shape.
The observational test of this template has been analyzed by using the WMAP7 data in \cite{Sefusatti:2012ye}, and by the Planck13 data in \cite{Planck:2013wtn}. We shall update the constraint using the Planck legacy release in Section \ref{sec:test}.

\item
The oscillatory collider signal arises when the massive scalar is in the principle series ($m>3H/2$). There on super-horizon scales the scalar particle decays and oscillates logarithmically in conformal time.
Through the exchange diagram, this produces a distinctive signature in the   scaling of the squeezed bispectrum \cite{Noumi:2012vr,Arkani-Hamed:2015bza,Lee:2016vti}
\begin{small}
\be \label{sq.col.}
\lim_{k_l\ll k_s} S^{\rm col.}  \propto  \(\frac{k_l}{k_s} \)^{\frac{1}{2}+i\mu} + c.c. \propto ~ \(\frac{k_l}{k_s} \)^{{1}/{2}}\cos\[\mu \ln\(\frac{k_l}{k_s}\)+ \delta\]~~~~
{\rm with}~ \mu=\sqrt{\frac{m^2}{H^2}-\frac{9}{4}}~.
\ee 
\end{small}From the oscillatory signal above, it becomes possible for us to do particle spectroscopy during inflation, as the frequency of the oscillation measures the mass of the intermediate  state. Away from the squeezed limit, in general for massive exchanges, it is very difficult to analytically derive the full shape functions of the scalar bispectrum.
While for the data analysis, it is important to have approximated bispectrum templates, like \eqref{qsf} for the quasi-single field scenario. In the following, we will tackle this problem with the help of the cosmological bootstrap. 
\end{itemize}

\paragraph{Bootstrapping scalar exchange templates}
One particular advantage of the bootstrap approach is to provide a powerful computational tool that allows us to gain the full analytical understanding of bispectrum shapes for any kinematics \cite{Arkani-Hamed:2018kmz,Pimentel:2022fsc,Jazayeri:2022kjy,Wang:2022eop}. Here we follow the approach developed in the {\it boostless cosmological collider bootstrap} \cite{Pimentel:2022fsc} and focus on deriving simpler analytical shape templates that can be easily applied in  data analysis. 

\begin{figure} 
   \centering
 \begin{subfigure}{0.4\textwidth}
      \includegraphics[height =4.6cm]{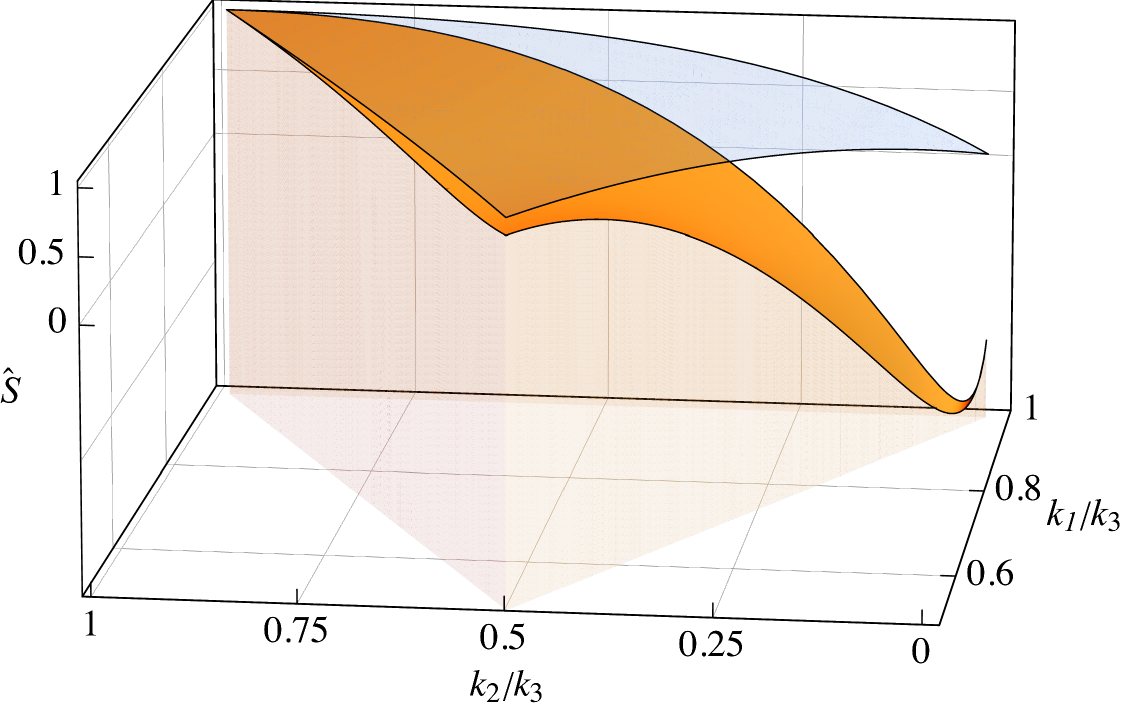} \caption{  $\mu=0.5$}
      \end{subfigure} \hspace{1cm}  
   \begin{subfigure}{0.4\textwidth}
      \includegraphics[height =4.6cm]{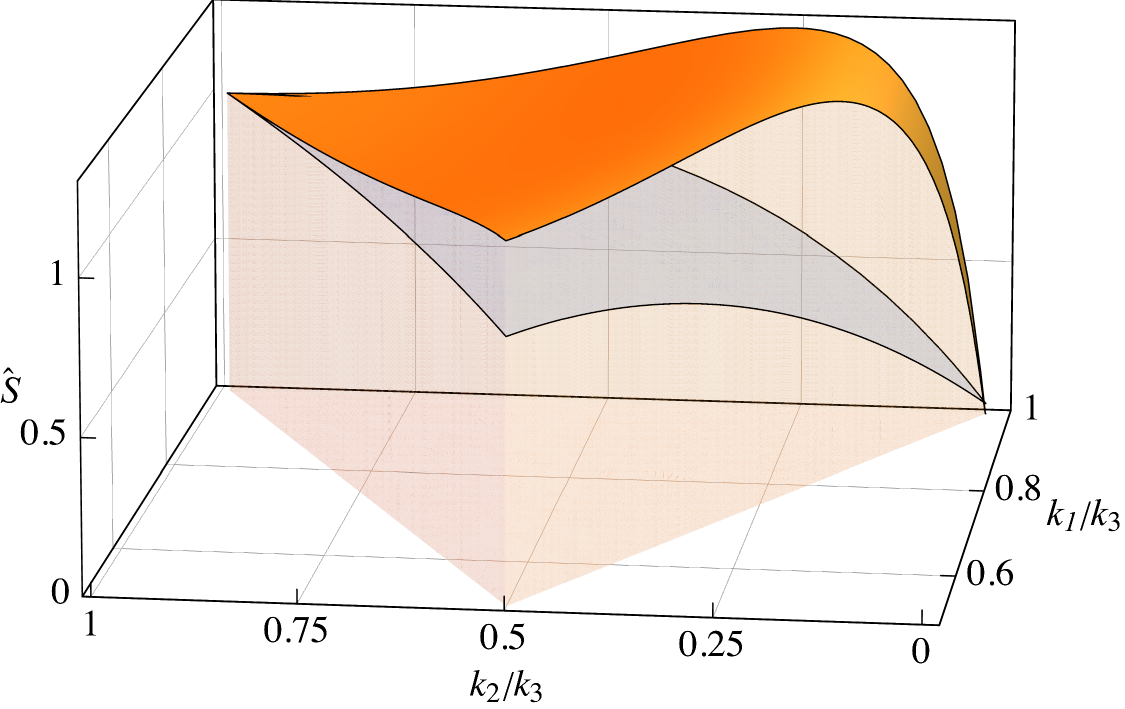} \caption{  $\mu=1$}
      \end{subfigure} \\ \vspace{0.5cm}
   \begin{subfigure}{0.4\textwidth}
      \includegraphics[height =4.6cm]{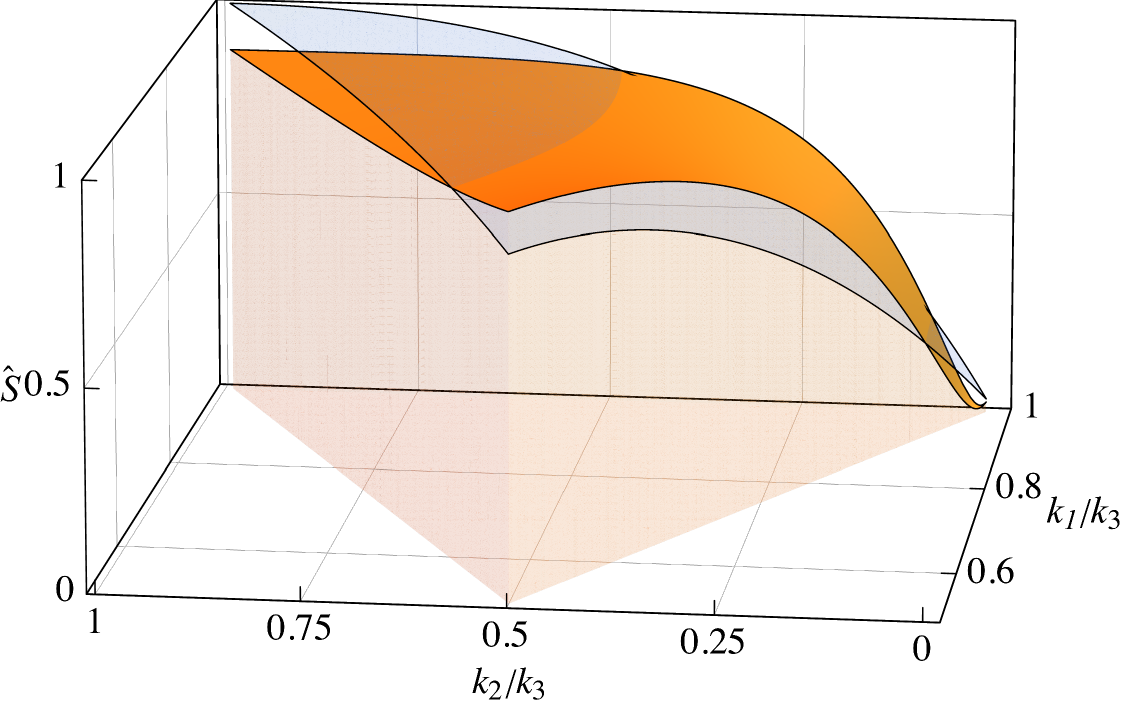} \caption{  $\mu=1.5$}
      \end{subfigure}  \hspace{1cm} 
   \begin{subfigure}{0.4\textwidth}
      \includegraphics[height =4.6cm]{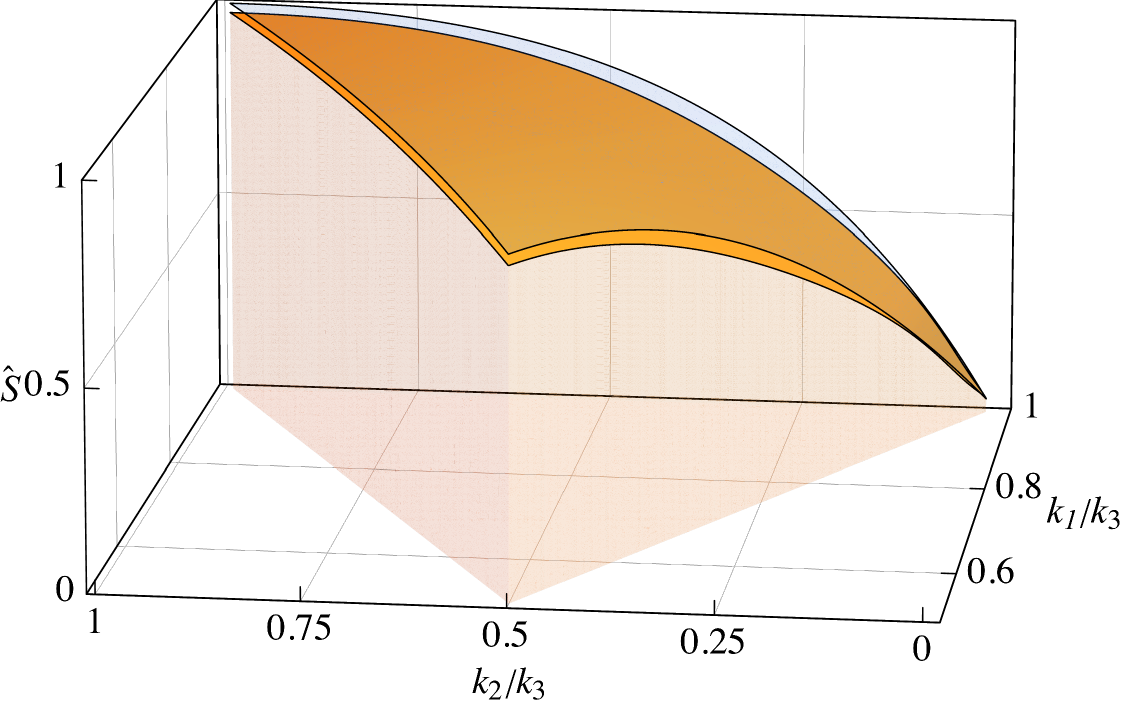} \caption{  $\mu=3$}
      \end{subfigure} 
      \caption{Equilateral-like shapes from massive scalar exchanges with different masses. The blue one is the standard equilateral shape generated by the single field EFT operator $\dot\phi^3$.} 
      \label{fig:shape-a}
\end{figure}

The starting point is an auxiliary correlator called the three-point scalar seed $\langle \vp^2\phi \rangle$, where $\vp$ is the conformally coupled scalar ($m^2=2H^2$), and $\phi$ is the massless scalar. 
This correlator is generated by the single exchange of a massive scalar $\s$, with one cubic vertex $\vp^2\s$ and one linear mixing $\dot\phi\s$. 
Then in the $s$-channel where $k_s=k_3$ is associated with the linear mixing leg, the three-point scalar seed can be written as a function of the following momentum ratio \footnote{In general, we can also introduce two  sound speed parameters for $\phi$ and $\sigma$ fields in the scalar seed. Here for simplicity we focus on the situation where both $\phi$ and $\sigma$ are canonical scalars, and leave the effects of nontrivial sound speed in Section \ref{sec:cs}.}
\be \label{I0}
\langle \vp^2\phi \rangle \sim \hat{\mathcal{I}}(u) ~,~~~~~~{\rm with}~ u\equiv \frac{ k_3}{k_1+k_2}~.
\ee
Without doing the nested time integrals, the bootstrap approach is to identify that from spacetime symmetries the scalar seed function $\hat{\mathcal{I}}(u) $ satisfies a boundary differential equation as follows
\be \label{bootstrapeq}
\(\Delta_u + \frac{m^2}{H^2} -2 \)  \hat{\mathcal{I}}(u) = \frac{u}{1 +  u}~,~~~~~~{\rm with}~ \Delta_u\equiv (1-u^2)u^2\partial_u^2 - 2u^3\partial_u ~.
\ee
By solving this equation with proper boundary conditions, we fully fix the analytical form of the scalar seed. In general, it contains two parts: one is the homogeneous solution of \eqref{bootstrapeq} given by hypergeometric functions $_2F_1$; the other is the inhomogeneous part given by a power series. Although this solution gives us the full analytical control of three-point scalar seed $\langle \vp^2\phi \rangle$, its form is rather complicated. 
For the purpose of deriving simple bispectrum templates for data analysis,
here we propose the following {\it approximated scalar seed} 
\begin{eBox2}
\be \label{Ia}
\hat{\mathcal{I}}_a(u) = \frac{u}{\beta} {(1+u)^{-\frac{\beta}{\beta+2}}}  - \frac{1}{2}\sum_{\pm}B_\pm\(\frac{u}{2}\)^{\frac{1}{2}\pm i\mu} + \tilde{f}(u)~,
\ee
\end{eBox2}
with
\be
\beta=\mu^2+\frac{1}{4} ~~~~~~{\rm and} ~~~~~~ B_\pm= \frac{\pi^{3/2}}{\cosh\pi\mu} \(1\mp\frac{i}{\sinh\pi\mu}\)
 \frac{\Gamma(\frac{1}{2}\pm i\mu)}{\Gamma(1\pm i\mu)}~.
\ee
The expression of $\tilde{f}(u)$ is given in \eqref{fu}.
The first two terms in \eqref{Ia} provide a good match with the exact result for  $\mu>1$, and there we can safely take $\tilde{f}(u)=0$.
The approximated scalar seed basically captures the two major properties of the exact solution of $\hat{\mathcal{I}}(u)$: i) the mass dependence of the non-squeezed kinematics configuration is approximated by the first term; ii) the second term gives us the collider signals in the squeezed limit. 
For $\mu>1$, we see the first term is more dominant, while the second term is exponentially suppressed and only contributes to the squeezed limit oscillations. Thus the final shape function looks more like the equilateral one, while on the top of it we have small wiggles around the corner of the squeezed configuration, as show in Figure \ref{fig:shape-a}.
We also notice that this simplification becomes less accurate when the scalar mass gets close to $3H/2$ (i.e. $0<\mu<1$), and we need to introduce the last term $\tilde{f}(u)$ to have a good approximation.
We leave the subtleties of this small mass regime and more details  of $\tilde{f}(u)$  in Appendix \ref{app:seed}.

Then the scalar bispectrum from massive exchange can be derived from the scalar seed using the weight shifting procedure
\be \label{phi3}
\langle \phi_{\bf k_1} \phi_{\bf k_2} \phi_{\bf k_3} \rangle' \sim \frac{1}{k_1^2k_2^2}  \mathcal{W}_{12} \[k_3 \hat{\mathcal{I}}(u) \]+ {\rm perms}~,
\ee
where $\mathcal{W}_{12}$ are differential operators associated with the form of the cubic vertices. The most general form of the weight-shifting operators for boost-breaking interactions is given in \cite{Pimentel:2022fsc}, using which we can derive the scalar exchange bispectrum from vertices with any number of derivatives.
Here we just focus on the following two simplest examples 
\bea
\mathcal{W}^{\dot\phi^2\sigma}_{12} &=& -  k_1k_2 \partial_{k_{12}}^2~,\\
\mathcal{W}^{(\partial_i\phi)^2\sigma}_{12} &=& -\frac{1}{2k_1k_2} (k_3^2 - k_1^2 - k_2^2) {\(1-k_1\partial_{k_1}\)\(1-k_2\partial_{k_2}\)}~.  \label{bbws}
\eea
Then substituting the approximated scalar seed  \eqref{Ia} into \eqref{phi3}, we find the simplified shape templates from massive scalar exchanges.
The result from the cubic vertex $\dot\phi^2\s$ is given by
\begin{eBox}\begin{small}\bea \label{scalarI}
S_{\rm col.}^{\rm scalar-I} &=& \frac{2k_1k_2k_3}{\beta k_T^2(k_1+k_2)} \[ 1+ \frac{4k_3}{ (\beta+2) (k_1+k_2)} + \frac{(\beta+4)k_3^2}{ (\beta+2)^2 (k_1+k_2)^2} \] \(\frac{k_T}{k_1+k_2}\)^{-\frac{\beta}{\beta+2}}  
+ \tilde{S}_I\\ 
&& + \frac{k_1k_2}{(k_1+k_2)^2} \sqrt{\frac{\pi^3\beta(\beta+2)}{\mu \sinh(2\pi \mu) }} \(\frac{k_3}{k_1+k_2}\)^{1/2} \cos\[\mu\log\(\frac{k_3}{2(k_1+k_2)}\) + \delta \] + 2~{\rm perms.}~, \nn
\eea    \end{small}     \label{eqn:template_SELM}
\end{eBox}
with 
\be
\delta = \arg \[ \Gamma\(\frac{5}{2}+i\mu\)\Gamma\(-i\mu\) (1+i\sinh\pi\mu) \]~.
\ee
Examples of this shape template are shown in Figure \ref{fig:shape-a}.
The second line  in \eqref{scalarI} represents the typical oscillatory signals in the squeezed limit, while
the first line gives us an equilateral-like shape that is deformed by the mass of the $\sigma$ field. 
There the second term $\tilde{S}_I\equiv -k_1k_2\partial_{k_1}\partial_{k_2}\tilde{f}$ is introduced for the small mass regime $\mu<1$ with its expression given in \eqref{ddfu}. 
In the heavy field regime $\mu>1$, the overall shape is dominated by
the non-oscillatory part, which resembles the standard equilateral shape, and simply reproduces $S\propto k_1k_2k_3/k_T^3$ for the $m\gg H$ limit.
On the top of it, the second line in \eqref{scalarI} gives the cosmological collider oscillations around the squeezed configuration. 
The situation becomes  different when the mass gets close to $3H/2$ (or $\mu\leq 1$). There, the oscillatory part and the non-oscillatory part are comparable with each other, and the overall shape could show large deviations from the equilateral one.

\begin{figure} 
   \centering
   \begin{subfigure}{0.4\textwidth}
      \includegraphics[height =4.6cm]{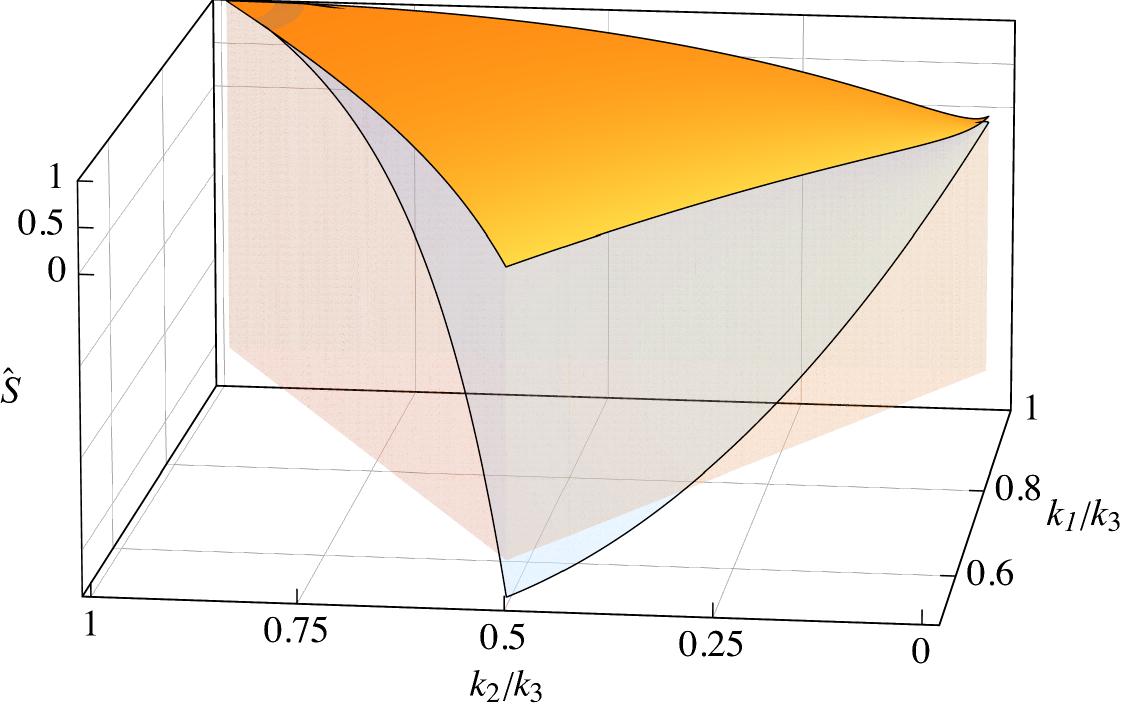} \caption{  $\mu=1$}
      \end{subfigure} \hspace{1cm}
   \begin{subfigure}{0.4\textwidth}
      \includegraphics[height =4.6cm]{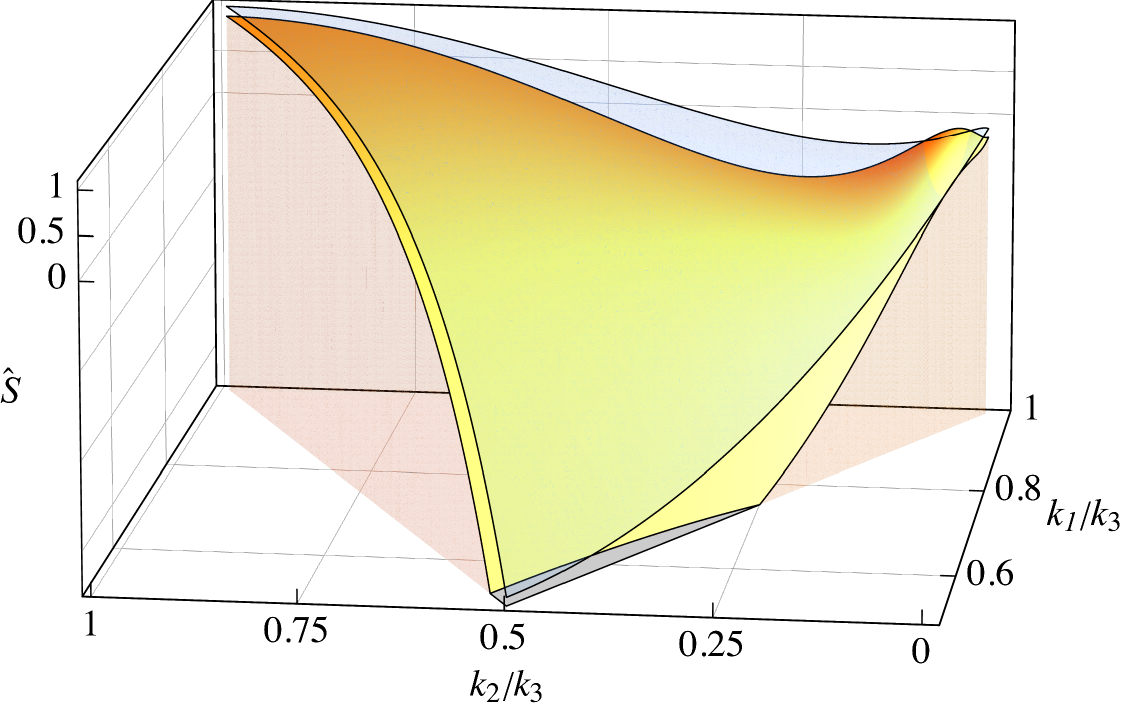} \caption{  $\mu=2$}
      \end{subfigure}  \\ \vspace{0.5cm}
       \begin{subfigure}{0.4\textwidth}
      \includegraphics[height =4.6cm]{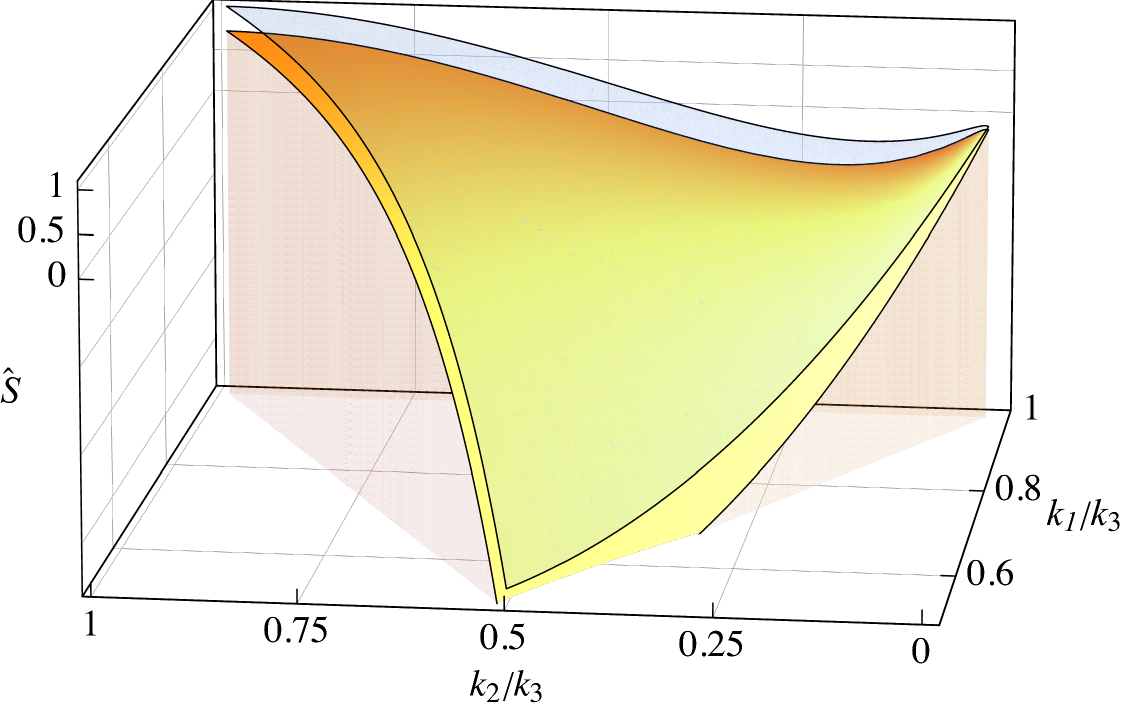} \caption{  $\mu=5$}
      \end{subfigure}  \hspace{1cm}
   \begin{subfigure}{0.4\textwidth}
      \includegraphics[height =4.6cm]{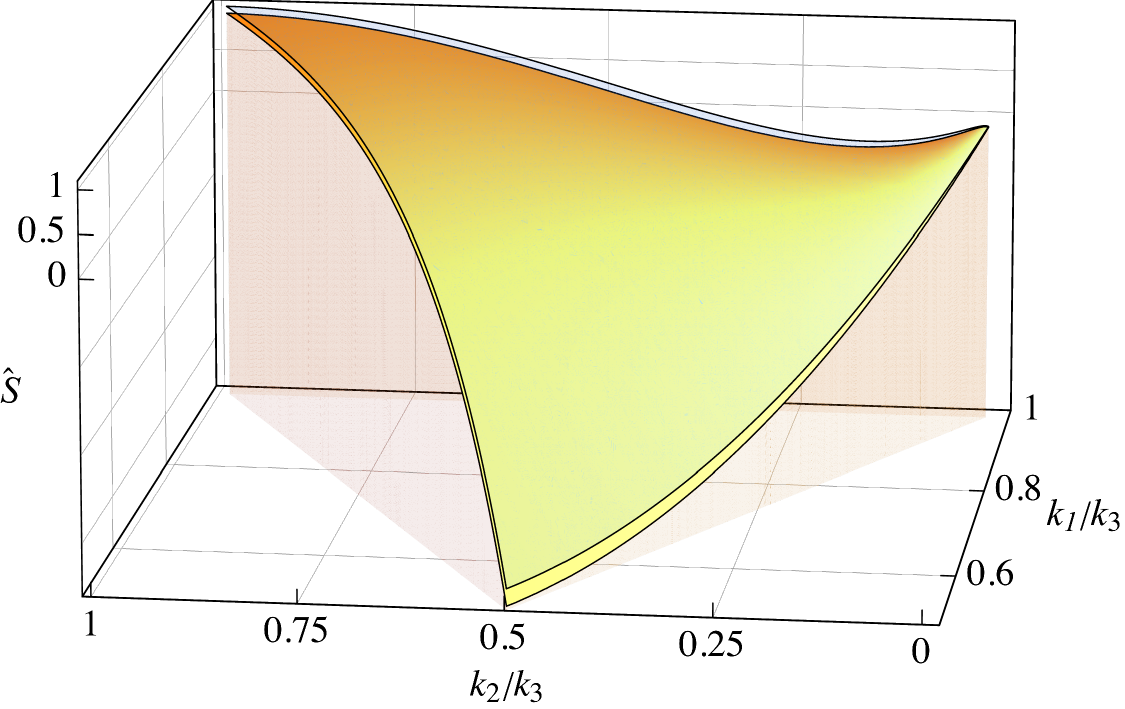} \caption{  $\mu=10$}
      \end{subfigure} 
      \caption{Orthogonal-like shapes from massive scalar exchanges with different masses. The blue one is the orthogonal shape generated by a linear combination of the single field EFT operators $\dot\phi^3$ and $\dot\phi(\partial_i\phi)^2$. } 
      \label{fig:ortho}
\end{figure}

Similarly, using the weight-shifting operator \eqref{bbws}, we derive the approximated template from another cubic vertex $(\partial_i\phi)^2\s$. 
\begin{small}\bea \label{scalarIIa}
S_{\rm col.}^{\rm II} &=& \frac{k_3(k_3^2-k_1^2-k_2^2)}{ \beta(k_1+k_2)^3} \[ 6-\frac{6\beta k_3}{(\beta+2)k_T} + \frac{2\beta(\beta+1)k_3^2}{ (\beta+2)^2 k_T^2} + 
\frac{k_1^2+k_2^2}{k_1 k_2}\(2-\frac{\beta k_3}{(\beta+2)k_T}\) \] \(\frac{k_T}{k_1+k_2}\)^{-\frac{\beta}{\beta+2}}  \nn
\\ 
&& + \frac{k_3^2-k_1^2-k_2^2}{k_1k_2} \(\frac{k_3}{k_1+k_2}\)^{1/2} \left\{ \sqrt{\frac{\pi^3(\beta+2)}{\mu \sinh(2\pi \mu) }} \cos\[\mu\log\(\frac{k_3}{2(k_1+k_2)}\) + \delta_1 \] \right.\nn\\
&& \left. + \frac{k_1k_2}{(k_1+k_2)^2}\sqrt{\frac{\pi^3\beta(\beta+2)}{\mu \sinh(2\pi \mu) }} \cos\[\mu\log\(\frac{k_3}{2(k_1+k_2)}\) + \delta_2 \]  \right\}  +   \tilde{S}_{II} + 2~{\rm perms.}~, 
\eea    \end{small}
with 
\bea
\delta_1 &=& \arg \[ \(\Gamma\(\frac{1}{2}+i\mu\) + \Gamma\(\frac{3}{2}+i\mu\)\)\Gamma\(-i\mu\) \(i+{\frac{1}{\sinh\pi\mu}}\) \]~,\\
\delta_2 &=& \arg \[ \Gamma\(\frac{5}{2}+i\mu\) \Gamma\(-i\mu\) \(i+{\frac{1}{\sinh\pi\mu}}\) \]~.
\eea
The expression of $\tilde{S}_{II}$ is given in \eqref{ddfu2}.
One can easily check that for $m\gg H$ the shape function goes back to the single field one from the EFT operator $\dot\phi(\partial_i\phi)^2$.
Meanwhile, recall that in single field inflation, although the two EFT shapes from $\dot\phi^3$ and $\dot\phi(\partial_i\phi)^2$ look similar with each other, a linear combination of the two leads to the new orthogonal shape, which is peaked both
at equilateral  and folded configurations \cite{Senatore:2009gt}.
Similarly here, for cosmological colliders, 
a linear combination of \eqref{scalarI} and \eqref{scalarIIa} gives us a new shape that resembles the orthogonal one but has squeezed-limit oscillations
\begin{eBox}
    \be
S_{\rm col.}^{\rm scalar-II} = - S_{\rm col.}^{\rm scalar-I} - 0.14 S_{\rm col.}^{\rm II} ~.    \label{scalarII}
\ee
\end{eBox}
Examples of this template are shown in Figure \ref{fig:ortho}.
Like in single field inflation, \eqref{scalarI} and \eqref{scalarII} provide a good basis to capture various possibilities of cosmological colliders from massive scalar exchanges.

The above analysis shows that, for masses with $\mu>1$, the cosmological collider signals from scalar exchange are small corrections to the standard equilateral or orthogonal shapes, which in general would serve as background signals of PNG. This is expected, since the oscillations in the squeezed limit are always suppressed by a Boltzmann factor $e^{-\pi\mu}$.  This is simply due to the fact that when  particles are heavier than the Hubble scale, it becomes more difficult to produce them during inflation.

As a final remark, we notice that the bootstrap analysis requires the full shapes from massive exchanges to have both the oscillatory and background parts.
While for the purpose of preparing simple shape templates, another possible approach is to consider the oscillatory parts only.
As we have seen, for $\mu>1$, the non-oscillatory parts mimic the bispectrum shapes from single field inflation. 
Thus, we may consider certain combinations of massive exchange and contact interactions to cancel the non-oscillatory  background and make the collider signal more manifest. For instance, in \eqref{scalarI}, this treatment would leave us with the second line there as the full template, which coincides with the {\it equilateral collider shape} in Section \ref{sec:cs}.
This approach may provide simpler and unified templates of cosmological colliders for observational tests, which we shall further examine in a follow-up work.  
In this paper, we will mainly take the full analytical shapes (``background + oscillations")  from the bootstrap computation for the data anlaysis.

\subsection{Shapes from spinning exchange}
\label{sec:spin}

Another typical signal of cosmological collider is the angular dependence from the exchange of spinning fields \cite{Arkani-Hamed:2015bza,Lee:2016vti,Arkani-Hamed:2018kmz,Pimentel:2022fsc} (see also \cite{Shiraishi:2013vja} for an early relevant work). This distinctive signature in the bispectrum can be seen as a consequence of cubic interactions between the inflaton and spinning particles. Two specific examples are
\be \label{spin-cubic}
\dot\phi \partial_{i_1i_2...i_s} \phi \sigma_{i_1i_2...i_s}~, ~~~~~~ \nabla_\mu\phi  \nabla^\mu \nabla_{\mu_1\mu_2...\mu_s}   \phi \sigma_{\mu_1\mu_2...\mu_s}~,
\ee
where the first one comes from the boost-breaking EFT \cite{Lee:2016vti,Pimentel:2022fsc}, and the second one is dS-invariant \cite{Arkani-Hamed:2015bza,Arkani-Hamed:2018kmz}. 
To have an exchange bispectrum, we also need linear mixings between the inflaton and spinning field, such as $ \partial_{i_1i_2...i_s} \phi \sigma_{i_1i_2...i_s}$. As a consequence, only the longitudinal mode of the spin-$s$ particle would contribute to the bispectrum, while  other components with nonzero helicities are  projected out by the linear mixing.
Therefore, we can simply assign the spin-$s$ field with a fixed tensor structure $\epsilon^L_{i_1i_2...i_s}$, which through the cubic vertices in \eqref{spin-cubic} leads to the universal angular-dependent profile for the bispectrum 
\be
k_2^{i_1}k_2^{i_2}...k_2^{i_s}\epsilon^L_{i_1i_2...i_s}({\bf k}_3) \sim P_s(\hat{\bf k}_2\cdot \hat{\bf k}_3)~,
\ee
where $P_s$ is the Legendre polynomial as a function of the angle between ${\bf k}_2$ and ${\bf k}_3$.
As a result, the squeezed bispectrum in \eqref{sq.col.} is generalized for spinning exchange
\be \label{sq.col.s}
\lim_{k_l\ll k_s} S^{{\rm spin}-s} \sim  P_s(\hat{\bf k}_l\cdot \hat{\bf k}_s) \(\frac{k_l}{k_s}\)^{1/2} \cos\[\mu \ln\(\frac{k_l}{k_s}\) +\delta\]~,
\ee
whose angular dependence measures the spin of the massive particles during inflation.
Away from the squeezed limit, the full shape expression for any kinematics is complicated as one can imagine. 
Again, in order to simplify the data analysis, we are motivated to look for approximated templates that have similar angular-dependent and oscillatory signals.

\paragraph{Heavy-spin exchange template} A simple approach is to consider the heavy field limit of the spin-$s$ particle $m_s\gg H$, where the intermediate state can be integrated out and we get a single field EFT. In this limit, the oscillations in the squeezed bispectrum become highly suppressed, but the angular dependence remains. We get the simple form of the heavy-spin exchange shape function in terms of rational polynomials \cite{MoradinezhadDizgah:2018ssw}
\begin{eBox}
\be
 S^{{\rm spin}-s}(k_1,k_2,k_3) =  
P_s(\hat{\bf k}_1\cdot \hat{\bf k}_3) \frac{k_2}{(k_1k_3)^{1-s} k_T^{2s+1}}\[ (2s-1)\( (k_1+k_3)k_T+2s k_1 k_3\) +k_T^2 \] + {\rm 5~ perms.}~,      \label{eqn:template_SE}
\ee
\end{eBox}
which can be seen as the equilateral shape with an angular-dependent profile.
However, we notice that
these shapes may also be predictions of single field inflation with higher-derivative interactions, such as $\dot\phi \partial_{i_1i_2...i_s} \phi  \hat\partial_{i_1i_2...i_s} \phi$, 
with $\hat\partial$ meaning that we only take the traceless part.

\begin{figure} 
   \centering
\begin{subfigure}{0.4\textwidth}
      \includegraphics[height =4.6cm]{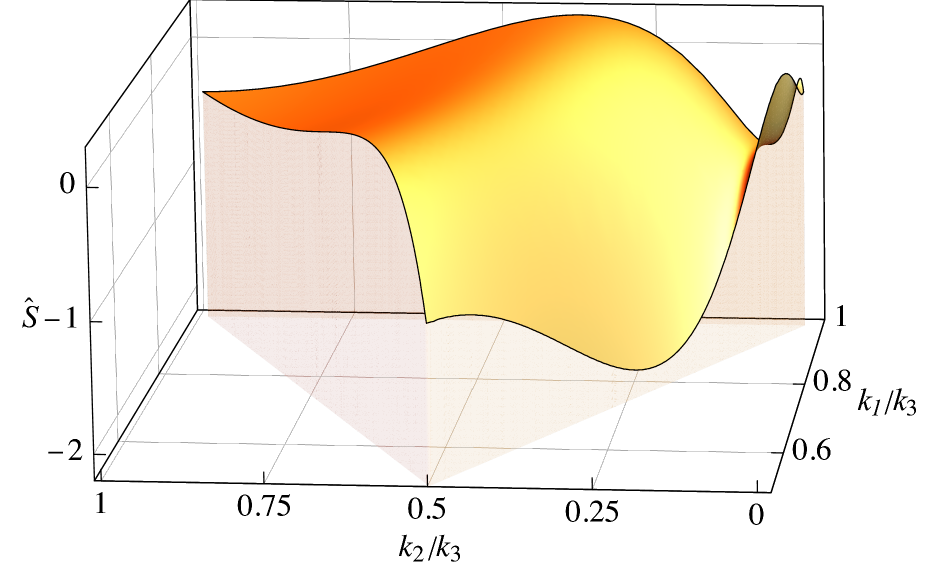} \caption{  $\mu=2$}
      \end{subfigure} \hspace{1cm}
\begin{subfigure}{0.4\textwidth}
      \includegraphics[height =4.6cm]{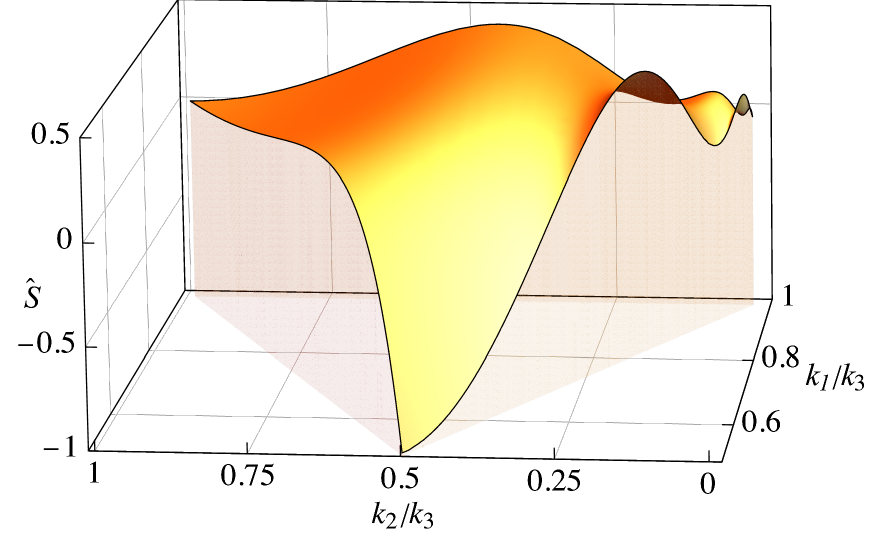} \caption{  $\mu=3$}
      \end{subfigure}  \\ \vspace{0.5cm}
       \begin{subfigure}{0.4\textwidth}
      \includegraphics[height =4.6cm]{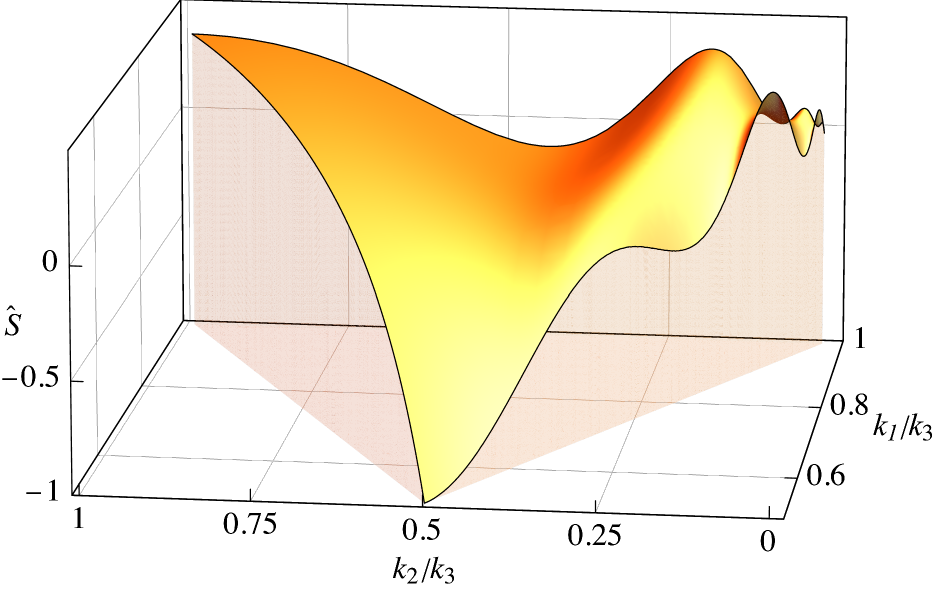} \caption{  $\mu=4$}
      \end{subfigure}  \hspace{1cm}
   \begin{subfigure}{0.4\textwidth}
      \includegraphics[height =4.6cm]{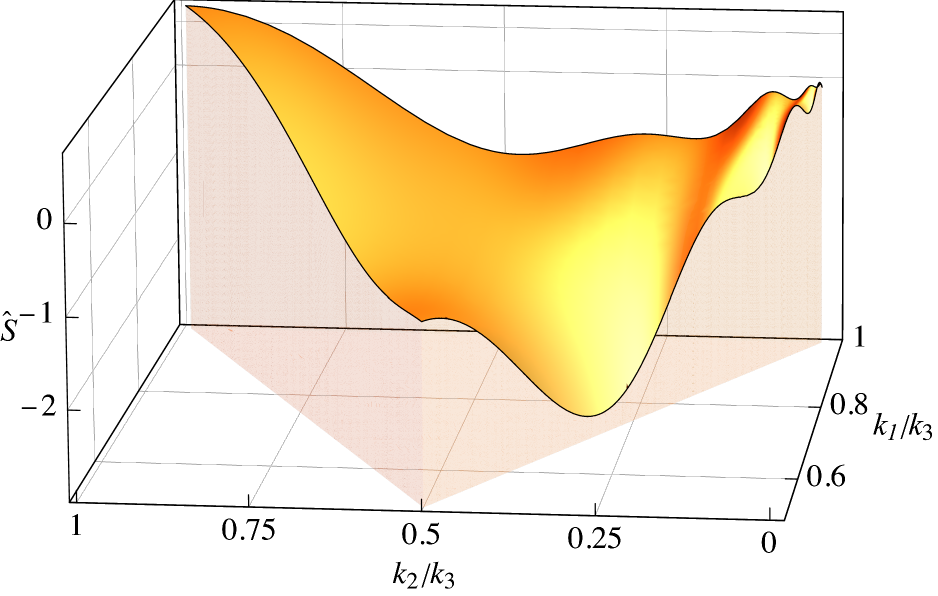} \caption{  $\mu=5$}
      \end{subfigure} 
      \caption{Angular-dependent shapes from spin-2 exchange template \eqref{spin2m} with different masses.} 
      \label{fig:spin2}
\end{figure}

\paragraph{Massive spin-2 exchange template}
To truly identify the distinctive signature of massive spinning particle, we still need to consider the $m\sim H$ regime and construct an approximated template for the shape functions there. As we have done for scalar exchanges, here we follow the bootstrap approach in Ref.~\cite{Pimentel:2022fsc} and  derive the template of  massive spin-2 exchange as an explicit example. The starting point is a generalized version of the approximated scalar seed $\hat{\mathcal{I}}_a^{(n)}(u)$, whose definition is given in \eqref{Ina}, and we leave more details in Appendix \ref{app:seed}.
Then the analytical templates of spin-$s$ exchanges can be obtained by using two types differential operators
\be \label{mspins}
S^{{\rm spin}-s}_{\rm col.}(k_1,k_2,k_3) =   P_s(\hat{\bf k}_2 \cdot \hat{\bf k}_3) {k_2^{s-1} k_3^{-s}} ~ \mathcal{W}^s_{12}  ~  \mathcal{D}^{(s)}_{23} ~ \left[ k_3 \hat{\mathcal{I}}^{(s)} \right]
+{\rm 5~ perms.}~,
\ee
where $\mathcal{W}^s_{12} = k_1\partial_{k_1}(1-k_2\partial_{k_2})$ is another weight-shifting operator and
 $\mathcal{D}^{(s)}_{23}$ is the spin-raising operator.
 For spin-2 exchanges, we have 
\be
\mathcal{D}_{23}^{(2)} \equiv  k_3^3 \[ \( \partial_{k_3}-\frac{2}{k_3}\)^2 \partial_{k_2} -\frac{1}{3c_s^2} \partial_{k_2}^3 \]
~,
\ee
which leads to the following bispectrum template
\begin{eBox}
\begin{small}\bea \label{spin2m}
S_{\rm col.}^{s=2} &=& P_2(\hat{\bf k}_2\cdot\hat{\bf k}_3)\frac{1991k_1k_2k_3^3}{ (\beta+6 )^2k_T^8} \[ (k_1+k_2)^2(k_1+6k_2)-(k_1+k_2)(7k_1+57k_2)k_3 + (-7k_1+8k_2)k_3^2+k_3^3 \]  \nn
\\ 
&& +P_2(\hat{\bf k}_2\cdot\hat{\bf k}_3)\frac{k_1k_2}{8(k_1+k_2)^3}  \sqrt{\frac{\pi^3\[1+\sinh^2(\pi\mu)\]}{2\mu \tanh(\pi \mu) }} \frac{\beta^2(\beta+2)^2}{\cosh^2(\pi\mu)} \(\frac{k_3}{k_1+k_2}\)^{1/2} \nn \\
&&\times \left\{ k_1\cos\[\mu\log\(\frac{k_3}{2(k_1+k_2)}\) + \delta_1^{s=2} \]
 +  \sqrt{\beta+12} k_2 \cos\[\mu\log\(\frac{k_3}{2(k_1+k_2)}\) + \delta_2^{s=2}  \]  \right\}   \nn
\\ 
&& -P_2(\hat{\bf k}_2\cdot\hat{\bf k}_3)\frac{k_1k_2}{6(k_1+k_2)^3}  \sqrt{\frac{\pi^3\beta(\beta+2)(\beta+6)(\beta+12)}{\mu \tanh(\pi \mu) }} \frac{\beta(\beta+2)}{\cosh(\pi\mu)} \(\frac{k_3}{2k_1+2k_2}\)^{5/2} \nn \\
&&\times \left\{ k_1\cos\[\mu\log\(\frac{k_3}{2(k_1+k_2)}\) + \delta_3^{s=2} \]
 +  \sqrt{\beta+30} k_2 \cos\[\mu\log\(\frac{k_3}{2(k_1+k_2)}\) + \delta_4^{s=2}  \]  \right\} \nn \\
 && +     5~{\rm perms.}~, 
\eea    \end{small}\end{eBox}
with 
\bea
\delta_1^{s=2} &=& \arg \[ \frac{\Gamma\(-i\mu\)}{\Gamma\(\frac{1}{2}-i\mu\)} \(1+{i\sinh\pi\mu}\) \]~,\\
\delta_2^{s=2} &=& \arg \[ \frac{\Gamma\(-i\mu\)}{\Gamma\(\frac{1}{2}-i\mu\)} \(1+{i\sinh\pi\mu}\) (7+2i\mu)\]~,\\
\delta_3^{s=2} &=& \arg \[  {\Gamma\(-i\mu\)}{\Gamma\(\frac{9}{2}+i\mu\)} \(1+{i\sinh\pi\mu}\) \]~,\\
\delta_4^{s=2} &=& \arg \[ {\Gamma\(-i\mu\)}{\Gamma\(\frac{9}{2}+i\mu\)} \(1+{i\sinh\pi\mu}\) \(\frac{11}{2}+i\mu\)\]~,
\eea
Although the expression is quite complicated, we can see that the first line here gives us the non-oscillatory part, while the rest is the collider signals in the squeezed limit.
Several examples are shown in Figure \ref{fig:spin2}. 
Interestingly, in addition to the angular dependence that changes the overall profile, the squeezed limit oscillations become more manifested in the spinning exchanges. This is because the non-oscillatory part of the template  is less dominant in \eqref{spin2m}, and the amplitudes of oscillations are comparably not-very-small  before getting exponentially suppressed for $\mu\gg 1$.

\subsection{Shapes with small sound speeds}
\label{sec:cs}

While the proposal of cosmological collider physics was first realized in (nearly) dS-invariant theories,  the size of the bispectrum there is in general suppressed by slow-roll parameters, which makes it extremely difficult to test these non-Gaussianity predictions. Then as in single field inflation, one of the most natural ways to enhance the signal is to apply the effective field theory approach, where dS boost symmetries can be broken and then small sound speeds of perturbations generically lead to detectably large non-Gaussianity.

More interestingly, the recent bootstrap analysis of these boost-breaking scenarios has identified new types of bispectrum shapes due to the presence of multiple sound speeds.
In general, it is possible for the inflaton $\phi$ and the mediator field $\s$ to have different sound speeds: $c_s$ and $c_\s$. 
Then with a large hierarchy between the two, intuitively speaking, the exchange process can probe multiple sound horizon crossings  before the $\sigma$ field decays into the inflaton. As a result, the inflaton leg with linear mixing may have  a different clock from the other inflaton legs. This in the end  modulates the non-Gaussian shapes of the primordial bispectrum.
In the following, we briefly review three types of new shape functions from cosmological colliders with nontrivial sound speed effects.

\paragraph{Equilateral collider}
In conventional scenarios,
the oscillatory signals mainly appear around the squeezed corner of the full shape.
This is because only there the oscillatory parts of the shape function dominates over the equilateral-type contributions, as we can see by comparing the scalings in \eqref{sq.col.} and \eqref{sq.col.s} with the one in \eqref{eq.scaling}.
In the full analytical results from the bootstrap analysis, the expansion parameter we use for the squeezed limit is $u\equiv k_3/(k_1+k_2)\ll 1$. For scenarios with two sound speeds, this parameter is extended to be $u\equiv c_\s k_3 /(c_sk_1+c_sk_2)$, which basically reflects the fact that the linear mixing leg $k_3$ inherits the sound speed of the intermediate $\s$ field.
For
$c_\s\ll c_s$, one interesting consequence is that we do not need to go to the actual squeezed triangle configuration to have $u\ll 1$, and therefore, the oscillatory collider signals would be able to appear even for equilateral configurations with $k_1\simeq k_2 \simeq k_3$. This is dubbed the {\it equilateral collider shape} \cite{Pimentel:2022fsc}, and one particular template from the cubic interaction $\dot\phi^2\s$ is given by
\begin{eBox}
\be \label{eqcol}
 S^{\rm eq.col.}(k_1,k_2,k_3) = \frac{k_1k_2}{(k_1+k_2)^2} \(\frac{k_3}{k_1+k_2}\)^{1/2}\cos\[\mu\log\(\frac{c_\s k_3}{2c_s(k_1+k_2)}\)+\delta^{\rm eq.col.} \] + {\rm 2~perms.}~.    
\ee
\end{eBox}
This shape function is similar with
the oscillatory part of the scalar exchange template I (the second line in \eqref{scalarI}), though here we have a nontrivial sound speed ratio $c_\s/c_s\ll 1$, which is important for the realization of this new shape.
Meanwhile, we also notice that there is a  degeneracy  between the sound speed ratio and the phase parameter $\delta^{\rm eq.col.}$. Thus for practical purpose, once we have shifted these oscillations outside of the squeezed limit, we can absorb  $\mu \log(c_\s/(2c_s))$ into the phase parameter, to further simplify the template expression.

In addition to the template in \eqref{eqcol},  there are certainly various possibilities for equilateral colliders, such as the ones from other types of scalar interactions, and also from spinning exchanges with angular-dependent profiles. In this work, we use this simplest example in the data analysis for demonstration, and we leave a more comprehensive investigation for future studies. 


\paragraph{Low-speed collider}
Once we turn on the hierarchy between the inflaton and the massive field sound speeds, another  possibility is $c_s\ll c_\s$. This is motivated by the consideration in the EFT of inflation that the boost-breaking operators  can lead to a very small sound speed for the inflaton fluctuations and naturally enhance the non-Gaussianity signal. Without loss of generality, here we can take $c_\s=1$. The bootstrap analysis of cosmological colliders with this setup has been performed in detail in Ref.~\cite{Jazayeri:2022kjy}, with a new bispectrum shape named as the {\it low-speed collider}.
Here we simply recap the physical picture and the non-Gaussian phenomenology.

For a two-field system during inflation, in addition to the sound horizon crossing of the inflaton at $k=a H/c_s$, there is also a mass horizon crossing for the massive field at $k=a m$.
Now we consider the massive field travels much faster than the inflaton fluctuations. When $m<H/c_s$, it becomes possible that these two types of  horizon crossings for different legs would happen at the coincident time. As a result, 
a resonance enhancement may happen around the mildly squeezed configuration
$k_l/k_s \simeq c_s m/H$, which generates a characteristic bump in the shape function.
In this situation, the oscillatory collider signal is further shifted deep into the squeezed limit, while the resonance peak in the mildly squeezed regime may become more dominant than the equilateral-type background part of the bispectrum.
A simple template for the low-speed collider was proposed in Ref.~\cite{Jazayeri:2023xcj}
\begin{eBox}
\be \label{lowcs}
 S^{{\rm LSC}}(k_1,k_2,k_3) =  S^{\rm eq}(k_1,k_2,k_3) +
 \frac{k_1^2}{3k_2k_3} \[ 1+ \alpha \(\frac{k_1^2}{3k_2k_3}\)^2\]^{-1} + {\rm 2~perms.}~,   
\ee
\end{eBox}
with $\alpha=c_sm/H$ and $S^{\rm eq}$ being the equilateral template used by WMAP and Planck. 
By varying the parameter $\alpha$, the location of the low-speed collider bump will be shifted. We present the data analysis in Section \ref{sec:results_LSC} and the constraints in Figure \ref{fig:LSC_constraints_1D}.
 
\paragraph{Multi-speed non-Gaussianity}
This type of new phenomenology is identified in the massless exchange diagram in Figure \ref{fig:png}c), and can be seen as one prediction of multi-field inflation that differs from the local shape.  
For multi-field models, the linear mixing $\dot\phi\s$ in general describes the usual isocurvature-to-curvature conversion process, and we expect infrared divergences in the massless exchange diagrams.  This provides a field-theoretical explanation  for the generation of local-type non-Gaussianities. 
However, if the linear mixing is given by higher derivative interaction, such as $\dot\phi\dot\sigma$ or $\partial^2\phi\s$, the infrared divergences disappear, and the mixed propagators can inherit the sound speed of the intermediate field.
As a result, for massless exchanges with multiple internal lines, each legs would have a different time of sound horizon crossing, controlled by their own sound speed. Similar with the low-speed collider, the resonance enhancement for the shape function may appear for arbitrary kinematic configurations depending on the sound speed ratios. More details about this new shape can be found in Ref.~\cite{Wang:2022eop}. Here we  use the following  template\footnote{See also \cite{Renaux-Petel:2011rmu} for an earlier work with a similar shape function but from a different setup.}
\begin{eBox}
\be \label{multics}
{S}^{{\rm multi}-c_s} (k_1,k_2,k_3)= \frac{k_1k_2k_3}{(c_1 k_1+ c_2 k_2 +c_3 k_3)^3} +  5~{\rm perms}~, 
\ee
\end{eBox}
where $c_{1,2,3}$ are different sound speed parameters. By tuning their ratios, we are changing the location of the peak in the bispectrum shape.
The data analysis on this template is presented in Section \ref{sec:results_EC_LSC_MS}, with results shown in Figure  \ref{fig:MS_significance}.

\section{CMB Bispectrum Estimator}
\label{sec:best}

In this section, we briefly review the current status of the CMB bispectrum analysis, and introduce the CMB-BEST python pipeline \cite{Sohn:2023fte}. This new high-resolution estimator uses a flexible set of modal bases for the primordial bispectra, and optimizes the algorithm to significantly improve the computational efficiency.
Particularly, it provides us with an ideal tool for testing PNG signals with complicated shape functions in the \textit{Planck} temperature and polarization data.

\subsection{CMB bispectrum statistics}

If the primordial fluctuations are non-Gaussian, the two-point function (power spectrum) no longer captures the full statistical information, and higher-order correlation functions can be significant. As the CMB anisotropies relate mostly linearly to the primordial perturbations, PNG is captured in higher-order statistics of the CMB.

\paragraph{CMB bispectrum} Given the observed CMB temperature anisotropies $a_{\ell m}$, the bispectrum statistic is constructed as the three-point function of the s as
\begin{align}
    \hat{B}^\ells_{m_1 m_2 m_3} \equiv a_{\ell_1 m_1} a_{\ell_2 m_2} a_{\ell_3 m_3} - \left[ \left< a_{\ell_1 m_1} a_{\ell_2 m_2} \right>_\mathrm{MC} a_{\ell_3 m_3} + (2\ \mathrm{ perms}) \right],    \label{eqn:observed_bispectrum}
\end{align}
where the terms in brackets account for the anisotropic sky coverage and masking effects, evaluated through ensemble averages (Monte-Carlo) from realistic simulations.

The expected value of $\hat{B}^\ells_{m_1 m_2 m_3}$ is zero if there is no PNG. Otherwise, given a primordial bispectrum shape $S(k_1,k_2,k_3;\boldsymbol{\theta})$ with some free parameters $\boldsymbol{\theta}$, 
\begin{align}
    \langle \hat{B}^\ells_{m_1 m_2 m_3} \rangle 
    &= \fNL\ \mathcal{G}^{\ells}_{m_1 m_2 m_3} \frac{144 P_\zeta^2}{5\pi^3}  \int r^2 dr dk_1 dk_2 dk_3 \; S (k_1,k_2,k_3;\boldsymbol{\theta}) \prod_{j=1}^{3} \left[ j_{\ell_j}(k_j r) T_{\ell_j} (k_j) \right] \\
    &= \fNL\ \mathcal{G}^{\ells}_{m_1 m_2 m_3} b_{\ells}(\boldsymbol{\theta}).
    \label{eqn:reduced_bispectrum}
\end{align}
The Gaunt integral $\mathcal{G}$ is a geometric factor\footnote{$\mathcal{G}^{\ells}_{m_1 m_2 m_3} \equiv \int d^2 \hat{\mathbf{n}}\ Y_{\ell_1 m_1}(\hat{\mathbf{n}}) Y_{\ell_2 m_2}(\hat{\mathbf{n}}) Y_{\ell_3 m_3}(\hat{\mathbf{n}}) $, where $Y_{\ell m}$s denote spherical harmonics.} that enforces the angular momentum conservation. The CMB transfer function $T_\ell (k)$ and spherical Bessel function $j_\ell(x)$ capture the evolution and projection of the primordial perturbations to the CMB anisotropies we observe today. The reduced bispectrum $b_{\ells}(\boldsymbol{\theta})$ contains all observable signatures of the theory in the CMB bispectrum, up to the overall amplitude $\fNL$. 

\paragraph{Likelihood} For CMB bispectrum analysis, we assume the weakly non-Gaussian regime where $\hat{B}^\ells_{m_1 m_2 m_3}$s are multivariate Gaussian with variance proportional to $C_{\ell_1} C_{\ell_2} C_{\ell_3}$ given by Wick's theorem. The CMB bispectrum likelihood is then given by
\begin{align}
    \mathcal{L}(\fNL, \boldsymbol{\theta}) &\propto \mathrm{exp} \left[ -\frac{1}{2} \sum_{\ell_j, m_j} \frac{1}{ 6\; C_{\ell_1} C_{\ell_2} C_{\ell_3}} \left| \hat{B}^\ells_{m_1 m_2 m_3} - \fNL\ \mathcal{G}^{\ells}_{m_1 m_2 m_3} b_{\ells}(\boldsymbol{\theta}) \right|^2 \right] \\
    &\equiv \mathrm{exp} \left[ (\mathrm{const.}) + \fNL P(\boldsymbol{\theta}) -\frac{1}{2} f^2_\mathrm{NL} F(\boldsymbol{\theta})  \right] .  
\end{align}
The function $P(\boldsymbol{\theta})$ and the Fisher information $F(\boldsymbol{\theta})$ are linear and quadratic in $b_\ells (\boldsymbol{\theta})$, respectively.

If the CMB bispectrum had been measured with high signal-to-noise, one could potentially perform a detailed analysis by varying both $\fNL$ and $\boldsymbol{\theta}$ to constrain the model space, analogously to the CMB power spectrum analysis where the $\Lambda$CDM model parameters are measured. However, we are in a regime with very low signal-to-noise so far, unfortunately. The CMB bispectrum analyses focus on fitting a single parameter $\fNL$.

For a given theoretical template $S(\boldsymbol{\theta})$, the best-fit $\fNL$ is given by the maximum likelihood estimator (MLE):
\begin{align}
    \hat{f}_\mathrm{NL}|_{\boldsymbol{\theta}}  = P(\boldsymbol{\theta}) / F(\boldsymbol{\theta}). \label{eqn:bispectrum_MLE}
\end{align}
This estimator is unbiased: $\langle \hat{f}_\mathrm{NL} \rangle =\fNL$ if the underlying model \eqref{eqn:reduced_bispectrum} is true. Furthermore, it is optimal, having the smallest expected variance amongst all unbiased estimators of $\fNL$ given $\boldsymbol{\theta}$: $\mathrm{Var}[\hat{f}_\mathrm{NL}|_{\boldsymbol{\theta}}]=F(\boldsymbol{\theta})^{-1}$, equal to the Cramer-Rao bound.

\paragraph{Normalization} The parametric form of \eqref{eqn:reduced_bispectrum} has a scaling redundancy: the theoretical bispectrum remains unchanged under $\fNL\rightarrow c\fNL$ and $S(k_1,k_2,k_3)\rightarrow S(k_1,k_2,k_3)/c$ for some constant $c$. This directly influences the $\fNL$ constraints quoted. Larger shape function amplitude gives smaller $\fNL$ estimates and seemingly tighter constraints. A shape function template's normalization therefore needs to be specified for clarity.

In this work, the templates have been normalized at the equilateral limit unless specified:
\begin{align}
    S (k_1, k_2, k_3) = 1 \quad \mathrm{at} \quad k_1=k_2=k_3=k_*.
\end{align}
The pivot value is fixed to $k_* \equiv 0.05\ \mathrm{Mpc}^{-1}$, although this would not affect the cosmological collider templates which are scale-invariant. This choice is equivalent to the \textit{Planck} normalization convention written in terms of the bispectrum of the gravitational potential $\Phi$:
\begin{align}
    B_\Phi^{\fNL = 1} (k_*, k_*, k_*) = 6 A^2,
\end{align}
where $A\equiv P_\Phi(k_*)$ is the primordial $\Phi$ power spectrum amplitude.

While this normalization convention provides a clear reference point to compare the detectability of various bispectrum shapes, it is not always the most fair one. In this convention, shapes with a relatively small equilateral limit and larger squeezed/flattened limits often have larger amplitudes, which lead to much tighter $\fNL$ constraints. In such cases, we direct the readers to focus on the significance $\fNL/\sigma(\fNL)$\textemdash the ratio of the estimated $\fNL$ and its uncertainty\textemdash as it is insensitive to the normalization.

\paragraph{Shape correlation} The shape and $\boldsymbol{\theta}$ are fixed in the $\fNL$ estimation shown in \eqref{eqn:bispectrum_MLE}. Independent studies of different shapes will hence provide different estimates of $\fNL$ from the observed CMB maps. However, the individual estimates are correlated. Suppose that two shapes $S^{(1)}$ and $S^{(2)}$ give two estimates $\hat{f}_\mathrm{NL}^{(1)}$ and $\hat{f}_\mathrm{NL}^{(2)}$. Their expected covariance is given by
\begin{align}
    \mathrm{Cov}\left[ \hat{f}_\mathrm{NL}^{(1)}, \hat{f}_\mathrm{NL}^{(2)} \right] = \frac{ \langle S^{(1)}, S^{(2)} \rangle_{\mathrm{CMB}} }{ \langle S^{(1)}, S^{(1)} \rangle_{\mathrm{CMB}} \ \langle S^{(2)}, S^{(2)} \rangle_{\mathrm{CMB}} },
\end{align}
where we defined the inner product $\langle \cdot \rangle_\mathrm{CMB}$ as a weighted dot product in the CMB bispectrum space:
\begin{align}
    \langle S^{(1)}, S^{(2)} \rangle_{\mathrm{CMB}} \equiv \sum_{\ell_j} \frac{h_\ells^2}{ 6\; C_{\ell_1} C_{\ell_2} C_{\ell_3}}  b^{(1)}_{\ells} b^{(2)}_{\ells}.
\end{align}
The reduced bispectrum $b^{(i)}_\ells$ relate to the shape $S^{(i)}$ through \eqref{eqn:reduced_bispectrum}. The geometric factor $h^2_\ells \equiv \sum_{m_j} \left| \mathcal{G}^\ells_{m_1 m_2 m_3} \right|^2$.  Using this notation, the Fisher information can be rewritten as $F=\left< S, S \right>_\mathrm{CMB}$.

We define the correlation (or `cosine') between two shapes as the Pearson correlation coefficient of their respective $\hat{f}_\mathrm{NL}$s:
\begin{align}
    \rho(S^{(1)},S^{(2)}) \equiv \mathrm{corr}\left[ \hat{f}_\mathrm{NL}^{(1)}, \hat{f}_\mathrm{NL}^{(2)} \right] = \frac{ \langle S^{(1)}, S^{(2)} \rangle_{\mathrm{CMB}} }{ \sqrt{ \langle S^{(1)}, S^{(1)} \rangle_{\mathrm{CMB}} \ \langle S^{(2)}, S^{(2)} \rangle_{\mathrm{CMB}} } }. \label{eqn:cosine_definition}
\end{align}
This value lies in $[-1,1]$ by construction. A strong correlation (+1) or anti-correlation (-1) between the two shapes implies that they probe similar forms of the CMB bispectrum, and their $\fNL$ estimates are highly correlated. Such shapes are difficult to distinguish from the CMB bispectrum analysis only. We also note that it is possible to have two very distinct shape functions with a strong correlation under this metric, since the projection from the primordial perturbations to the CMB anisotropies can wash out the differences.

\begin{figure} 
    \centering
    \includegraphics[]{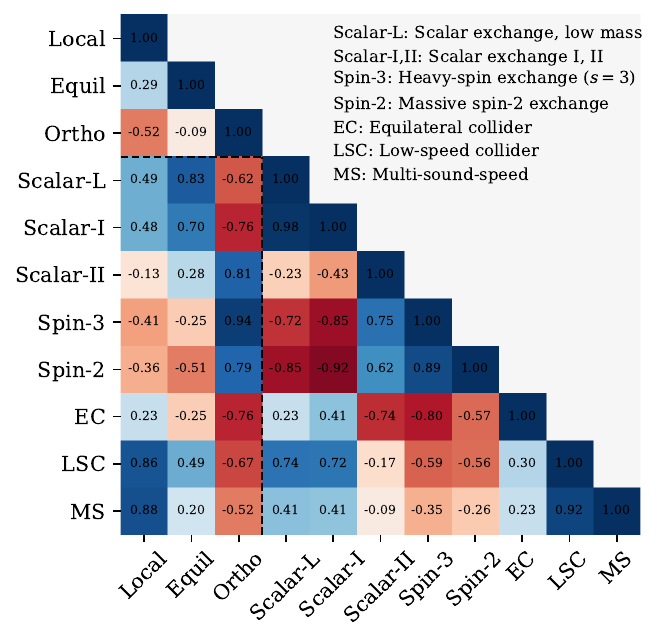}
    \caption{Correlation between the cosmological collider templates studied in this work, together with the standard local, equilateral and orthogonal templates. The correlation (defined as \eqref{eqn:cosine_definition}) measures the similarity of their observable CMB bispectrum predictions. Free parameters in each template have been fixed to their best-fit values for this plot. Here we also summarize our acronyms for templates.}
    \label{fig:all_stars_correlation}
\end{figure}

Figure \ref{fig:all_stars_correlation} shows the correlation between cosmological colliders templates and the standard local, equilateral and orthogonal templates. Note that the templates were evaluated at their best-fit parameters, {and any change in the parameters can have a significant impact on the correlation values shown here};  
refer to Section \ref{sec:test} for detailed analysis. We also take this chance to introduce the acronyms of the new templates which will be used later. Many of the templates studied in this work are not strongly correlated with any of the standard templates.

\subsection{CMB bispectrum estimator}

\paragraph{Mode expansion} Direct numerical computation of  \eqref{eqn:bispectrum_MLE} is prohibitively expensive due to the number of terms appearing in the sum: $O(\ell_\mathrm{max}^5)$ for $\ell_\mathrm{max}\lesssim 2500$ for \textit{Planck}. Existing formalisms utilise the separability of the high-dimensional integral \cite{Komatsu:2003iq,Creminelli2006limits,Yadav2007,Senatore:2009gt,Smith:2006ud,Fergusson:2009nv,Fergusson:2010dm,Fergusson:2014gea} or binning to reduce the computational complexity \cite{Bucher_2010,Bucher:2015ura}. In CMB-BEST, the (primordial) shape function is expanded in terms of a separable basis, so that
\begin{align}
    S(\boldsymbol{\theta})(k_1, k_2, k_3) = \sum_{n \leftrightarrow (p_1,p_2,p_3)} \alpha_n (\boldsymbol{\theta}) \; q_{p_1}(k_1) q_{p_2}(k_2) q_{p_3}(k_3), \label{eqn:basis_expansion}
\end{align}
where the one-dimensional basis functions $q_p(k)$ are Legendre polynomials of order up to $p_\mathrm{max} = 30$, forming a three-dimensional basis of size $\sim 5000$ up to symmetry. The basis function choices are flexible in principle, but we have found that our Legendre basis accurately covers the range of templates studied in this work.

Thanks to the separability of the terms appearing in \eqref{eqn:basis_expansion}, the three-dimensional integrals $\int dk_1 dk_2 dk_3$ appearing in \eqref{eqn:reduced_bispectrum} split into three one-dimensional integrals, dramatically reducing the computational complexity. Under the CMB-BEST formalism, only the expansion \eqref{eqn:basis_expansion} in the primordial space is required, and the CMB bispectrum likelihood is given by
\begin{align}
    \log \mathcal{L}(\fNL, \boldsymbol{\theta}) = \log\mathcal{L}_0  + \fNL \sum_n \beta_n \alpha_n (\boldsymbol{\theta})  - \frac{1}{2} f^2_\mathrm{NL} \sum_{n,n'} \gamma_{n n'} \alpha_n (\boldsymbol{\theta}) \alpha_{n'} (\boldsymbol{\theta}). \label{eqn:bispectrum_loglikelihood}
\end{align}
Here, $\mathcal{L}_0$ is a constant corresponding to the likelihood of $\fNL=0$, or vanishing bispectrum. The quantities $\beta_n$ and $\gamma_{n n'}$ only need to be computed once per basis and CMB map. The ones corresponding to the Legendre basis and the Planck 2018 CMB map (both temperature-only and with polarization) are provided together with the public CMB-BEST code. The best-fit $\fNL$ can be written as
\begin{align}
    \hat{f}_\mathrm{NL}|_{\boldsymbol{\theta}}  = \frac{ \sum_n \beta_n \alpha_n (\boldsymbol{\theta}) }{ \sum_{n,n'} \gamma_{n n'} \alpha_n (\boldsymbol{\theta}) \alpha_{n'} (\boldsymbol{\theta}) }. \label{eqn:bispectrum_MLE_alpha}
\end{align}
We use 160 \textit{Planck} full focal plane simulations with Gaussian initial conditions for two purposes: 1) to estimate the ensemble average in the linear term of \eqref{eqn:observed_bispectrum} and 2) to compute the best-fit $\fNL$s on each map to evaluate the estimation uncertainty: $\sigma(\hat{f}_\mathrm{NL})$.

\paragraph{Significance} We define the significance, or the signal-to-noise ratio, as the ratio of the estimated $\fNL$ and its estimation uncertainty:
\begin{align}
    \sigmasnr|_{\boldsymbol{\theta}} \equiv \left| \frac{ \hat{f}_\mathrm{NL}|_{\boldsymbol{\theta}} }{ \sigma(\hat{f}_\mathrm{NL}|_{\boldsymbol{\theta}}). } \right|.
\end{align}
The larger the $\sigmasnr$, the more significant the evidence towards a non-zero $\fNL$ for the given template. Formally put, under the null hypothesis of vanishing PNG ($H_0:\ \fNL=0$), the alternative hypothesis of  a non-zero bispectrum signal ($H_1:\ \fNL \neq 0$) can be tested. For a fixed value of $\theta$, the corresponding $p$-value is given by $p=1-\mathrm{erf}(\sigmasnr/\sqrt{2})$ under the likelihood \eqref{eqn:bispectrum_loglikelihood}.  A test with a significance level of $\alpha = 0.05$, for example, rejects $H_0$ whenever $\sigmasnr > 2$.

It is common to test multiple templates or a template with varying parameters $\boldsymbol{\theta}$ simultaneously. In such cases, one obtains a set of $\sigmasnr |_{\boldsymbol{\theta}}$, which provides a higher probability of finding at least one large number by chance. One may make a false detection by simply `looking elsewhere' until a large significance is found. Such look-elsewhere effect must be accounted for in any extensive parametric searches for PNG. 

We follow the ideas presented in \cite{Fergusson:2015le} and consider the statistical distribution of the raw maximum significance $\sigma_\mathrm{max} \equiv \mathrm{max}_{\boldsymbol{\theta}} \left[ \sigmasnr|_{\boldsymbol{\theta}} \right]$. This distribution is then used to compute the look-elsewhere-adjusted significance $\tilde{\sigma}_\mathrm{SNR}$. More precisely, given the raw maximum significance $\sigma_\mathrm{max}$, the $p$-value of the joint hypothesis test is given by
\begin{align}
    p(\sigma_\mathrm{max}) = 1 - \mathrm{Prob}\left[ \sigmasnr|_{\boldsymbol{\theta}} \le \sigma_\mathrm{max}\ \mathrm{for\ all}\ \boldsymbol{\theta} \right],
\end{align}
which gives the adjusted significance level
\begin{align}
    \tilde{\sigma}_{\mathrm{SNR}} \equiv \sqrt{2}\, {\mathrm{erf}}^{-1} \left(  1 - p(\sigma_{\mathrm{max}}) \right),   \label{eqn:adjusted_SNR_from_p}
\end{align}
where $\mathrm{erf}^{-1}$ is the inverse error function.

In practice, we compute $\tilde{\sigma}_{\mathrm{SNR}}$ for a template scan with $\boldsymbol{\theta} = \boldsymbol{\theta}_i $ for $i=1,\cdots,N$  as follows. Under the null hypothesis of vanishing bispectrum, the theoretical correlation between significances is given by \eqref{eqn:cosine_definition}: $C_{ij} \equiv \rho\left(S(\boldsymbol{\theta}_i), S(\boldsymbol{\theta}_j) \right) =  F_{ij} / \sqrt{F_{ii} F_{jj}}$ (no summation assumed), where the Fisher matrix $F_{ij} = \sum_{n,n'} \gamma_{nn'} \alpha_n(\boldsymbol{\theta}_i) \alpha_{n'}(\boldsymbol{\theta}_j) $ follows from the likelihood \eqref{eqn:bispectrum_loglikelihood}. We create $10^5$ realisations of the multivariate normal distribution $\mathcal{N}(\mathbf{0},C)$, and compute the maximum significance for each realisation. Next, we find the fraction of realisations that have their maximum significance greater than the observational value $\sigma_\mathrm{max}$. This fraction accurately estimates $p(\sigma_\mathrm{max})$, which can then be used to compute $\tilde{\sigma}_{\mathrm{SNR}}$ through \eqref{eqn:adjusted_SNR_from_p}.

\section{The Search for Collider Signals in the Planck Data}
\label{sec:test}

In this section, we apply the CMB-BEST for data analysis and present constraints on cosmological colliders based on various templates introduced in Section \ref{sec:shape}. All constraints are based on \textit{Planck} 2018 temperature and polarisation data.

The main objective of the data analysis shown here is twofold. First, we provide explicit bounds on the $\fNL$ for various templates. This serves as a reference for theoretical model-building and can rule out models that predict larger amplitudes. Second, we search for signatures of cosmological colliders by studying the significance of non-zero $\fNL$ for various templates. Statistically significant detection would have important implications for early universe physics.

The shape templates are normalized so that $S(k_*,k_*,k_*)=1$ unless explicitly specified, as discussed in Section \ref{sec:best}. The uncertainties on the $\fNL$ constraints are computed from 160 \textit{Planck} simulated maps with Gaussian initial conditions ($\fNL=0$). All $\fNL$ constraints quoted explicitly are at $68\%$ confidence level.


\subsection{Oscillations from massive scalars} \label{sec:results_QSF_SELM_OC}

We start with the analysis of three bispectrum templates from massive scalar exchanges proposed in Section \ref{sec:scalar}.

\begin{figure} [t]
    \centering
    \includegraphics[width=0.46\textwidth]{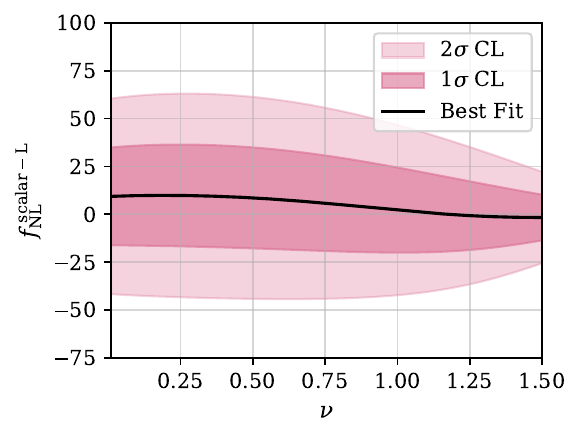}
~~\includegraphics[width=0.46\textwidth]{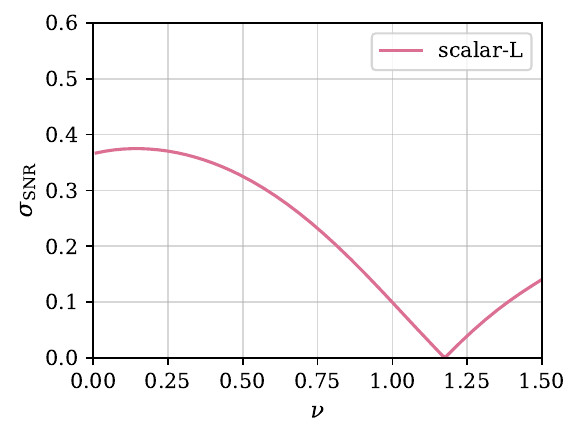}
    \caption{\textit{Left}: Constraints on the quasi-single field inflation template \eqref{qsf} for allowed values of the index ${0<\nu< 3/2}$. The black solid line indicates the best-fit $\fNL$ value for each value of $\nu$ considered independently. The 68\% ($1\sigma$) and 95\% ($2\sigma$) confidence levels are shown. \textit{Right}: The raw significance $\sigmasnr$ is shown as a function of $\nu$, maxmised at $\nu=0.14$ with maximum value $0.37$.
    }
    \label{fig:QSF_constraints_1D}
\end{figure}

\paragraph{Light scalar exchange} \label{sec:results_QSF}
The non-oscillatory template \eqref{qsf} is generated from massive scalars in complementary series, and has one free mass parameter $\nu\in [0,3/2]$ in addition to the size of the bispectrum $\fnl^{\rm scalar-L}$. The constraints on this template at each value of $\nu$ are shown in Figure \ref{fig:QSF_constraints_1D}. The best-fit values of $\fNL$ are shown together with $1\sigma$ and $2\sigma$ confidence levels.

Note that this template has been constrained previously in the \textit{Planck} 2013 analysis \cite{Planck:2013wtn}, and our constraints are consistent with these results. However, we do not perform the same Monte-Carlo Markov Chain (MCMC) analysis as shown in \cite{Planck:2013wtn} and refrain from setting a prior distribution on $\nu$. Instead, the independent constraints for each $\nu$ are summarised in Figure \ref{fig:QSF_constraints_1D}.

The constraints are given by $9 \pm 26$ at $\nu=0$ and $-2 \pm 12$ at $\nu=1.5$ at $68\%$ confidence level (CL hereafter). The error bars are tighter for larger $\nu$ because our normalization convention fixes the equilateral limit; the shapes with larger $\nu$ have larger squeezed-limit contributions to the bispectrum, and larger bispectrum templates get tighter constraints on their amplitudes $\fNL$.

Overall, having $f_\mathrm{NL}^\mathrm{scalar-L}=0$ is entirely consistent with the data. The righthand plot of Figure \ref{fig:QSF_constraints_1D} shows the significance (signal-to-noise) of $\fnl^{\rm scalar-L}$ as a function of $\nu$, which takes the maximum value of $0.37$ at $\nu=0.14$. This value is rather small. To put this into context, we studied the theoretical distribution of the maximum significance ($0\le \nu \le 1.5$) statistic. The maximum significance is greater than or equal to 0.37 about 90\% of the time (or less than 0.37 about 10\% of the time). Our results are fully consistent with having Gaussian initial conditions.

\paragraph{Massive scalar exchange} \label{sec:results_SELM_OC}

Two templates \eqref{scalarI} and \eqref{scalarII} correspond to the equilateral and orthogonal types of massive scalar bispectra, respectively. Both shapes contain a single free mass parameter $\mu$. For $\mu\gg1$, they reproduce the standard equilateral and orthogonal shapes because the oscillations are suppressed by the Boltzmann factor $e^{-\pi\mu}$. In the bispectrum data analysis, we scan through $0<\mu<6$.

\begin{figure}
    \centering
    \includegraphics[width=0.47\textwidth]{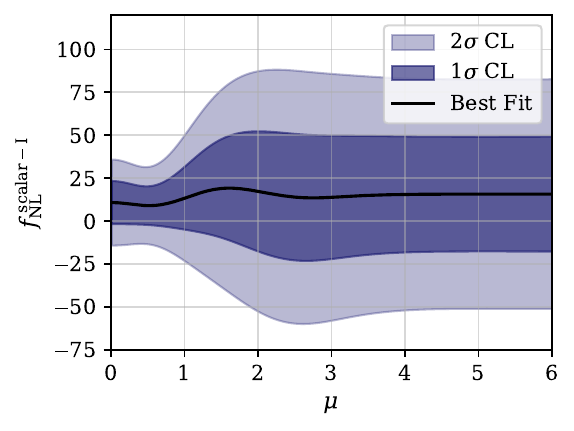}
    \includegraphics[width=0.47\textwidth]{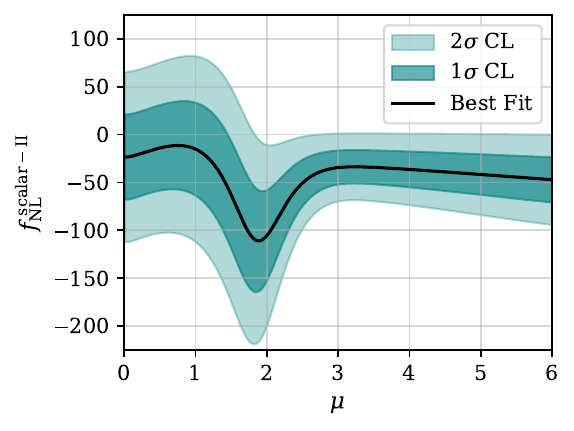}
    \caption{Constraints on the scalar exchange templates \eqref{scalarI} (Left) and \eqref{scalarII} (Right) for the index $\mu$ varied from $0$ to $6$. The black solid line indicates the best-fit $\fNL$ value for each value of $\mu$ considered independently, and the 68\% and 95\% confidence levels are shown in shaded colours. The amplitude $\fNL$ is constrained at the $\sigma(\fNL)\sim 10-50$ level. }
    \label{fig:scalar_constraints_1D}
\end{figure}

Figure \ref{fig:scalar_constraints_1D} shows the constraints for the two scalar exchange templates. On the left-hand side is the first template \eqref{scalarI}. As before, the best-fit value of $\fNL$ and the $1\sigma$ and $2\sigma$ error bounds are shown for each $\mu$. The constraints are given by $\fNL=11\pm 13$ at $\mu=0$, widening to $\fNL = 13 \pm 18$ at $\mu=1$, and converging around $\fNL = 16 \pm 33$ for $\mu>5$, which simply agrees with the constraint on the equilateral shape. Vanishing PNG ($\fNL=0$) lies within $1\sigma$ of the contours everywhere; the data is fully consistent with zero bispectrum amplitude for this template.

Shown on the right-hand side of Figure \ref{fig:scalar_constraints_1D} is the scalar exchange template II \eqref{scalarII} constraints. The shape function here approaches the orthogonal template when we increase the mass. The constraints are $\fNL=-24\pm 45$ at $\mu=0$, shifted to $\fNL = -15.7 \pm 49$ at $\mu=1$. The lowest best-fit value is reached for  $\fNL=-111\pm 53$ at $\mu=1.89$, and slowly changes to $\fNL = -59 \pm 29$ at $\mu=6$. Unlike the equilateral-type template earlier, $\fNL=0$ is outside the $2\sigma$ contour for some parts of the parameter space around $\mu=2.1$

\begin{figure}
    \centering
    \includegraphics[width=0.46\textwidth]{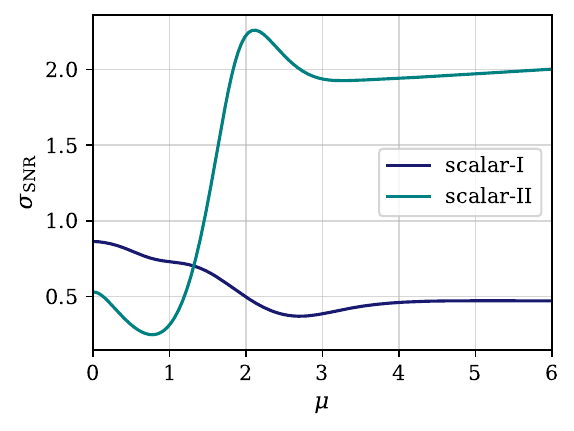}~~
    \includegraphics[width=0.45\textwidth]{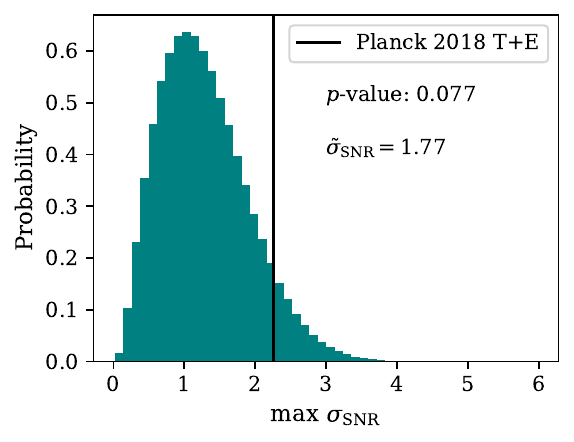}
    \caption{Significance of the best-fit $\fNL$ values for massive scalar exchange templates \eqref{scalarI} and \eqref{scalarII}. \textit{Left}: The raw significance $\sigmasnr$ is shown as a function of $\mu$ for the two templates. \textit{Right}: The scalar-II template's maximum significance of 2.3 at $\mu=2.1$ is compared with the theoretical distribution of the statistic. We find the significance as big as 2.3  $7.8\%$ of the time and the adjusted significance of 1.8.  }
    \label{fig:scalar_significance}
\end{figure}

Figure \ref{fig:scalar_significance} shows the significance of signals for the two scalar exchange templates. For the first template, the significance is less than 1 for all values of $\mu$ and is statistically consistent with random fluctuations around $0$. We obtain the maximum significance of $0.86$ about $50\%$ of the time in a Universe with Gaussian initial conditions, corresponding to the look-elsewhere-effect-adjusted significance of $0.67$.

Meanwhile, the orthogonal-type template is more favoured by the \textit{Planck} CMB data. The maximum significance of $2.3$ is obtained at $\mu=2.13$, which is adjusted to give $\tilde{\sigma}_\mathrm{SNR}=1.8$.  This is not sufficiently strong statistically to reject $\fNL=0$, but we note that this is the \textit{largest} adjusted significance found for the cosmological collider shapes so far.

\subsection{Angular dependence from spinning particles}
\label{sec:results_SE_SO}

Next, we turn to analyze the two types of bispectrum templates with angular dependence proposed in Section \ref{sec:spin}.

\begin{figure}
    \centering
    \includegraphics[]{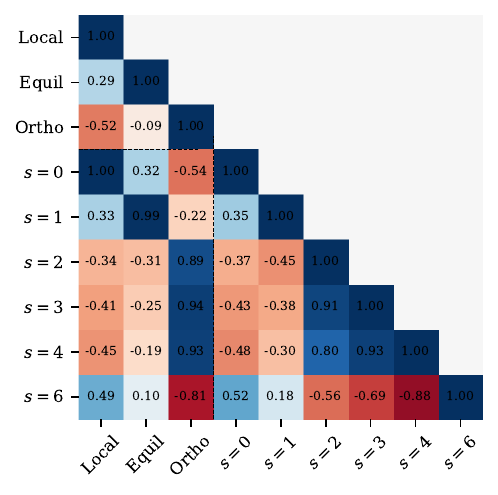}
    \caption{Correlation between the standard local, equilateral and orthogonal templates and the heavy-spin exchange templates \eqref{eqn:template_SE}. The correlation is defined in \eqref{eqn:cosine_definition} and measures the similarity of their observable CMB bispectrum predictions, with $1$ and $-1$ for perfect correlation and anti-correlation, respectively. }
    \label{fig:SE_correlations}
\end{figure}

\paragraph{Heavy-spin exchange}
\label{sec:results_SE}

The template \eqref{eqn:template_SE} probes the characteristic angular dependence from spinning particles with $m_s\gg H$, which can be seen as the predictions of single field interactions after integrating out a heavy spin-$s$ field. The spin $s$ enters the template as the order of the Legendre polynomial, and higher spins lead to more complex angular behaviours. Figure \ref{fig:SE_correlations} shows the correlation between the heavy-spin exchange templates and the standard local, equilateral and orthogonal templates. Spin $0$ and $1$ cases are indistinguishable from the local and equilateral templates with tiny differences due to small spectral tilt, as expected from the formula \eqref{eqn:template_SE}. Spins $2$-$4$ show strong to moderate correlations with the orthogonal template and each other. Spin 6 probes a relatively new shape different from the standard templates.

\begin{table}[h]
    \centering
    \renewcommand{\arraystretch}{1.5} 

    \begin{tabular}{lccc}
    \toprule
	   Spin & $\fNL$ constraint & Significance
 \\
    \midrule
       0 & $-1.5 \pm 5.8$ & 0.26   \\
       1 & $-8 \pm 45$ & 0.18     \\
       2 & $-18 \pm 10$ & 1.8   \\
       3 & $-59 \pm 32$ & 1.8   \\
       4 & $-17 \pm 16$ & 1.1   \\
       6 & $-1.9 \pm 6.7 $ & 0.28   \\
       
    \bottomrule
    \end{tabular}
    \caption{Constraints on the heavy-spin exchange template \eqref{eqn:template_SE}, for different values of spin. Errors are quoted at $68\%$ CL. The significance $\sigmasnr$ for each template is shown. }
    \label{tab:SE_constraints}
\end{table}

The constraints on $\fNL$ are shown in Table \ref{tab:SE_constraints}. We limit ourselves to smaller and even spins based on Occam's razor. We note that this choice is somewhat arbitrary and one can in principle study larger values of $s$.
The $\fNL$ errors for spins $0$ and $1$ are close to the ones for the local and equilateral templates, but the errors tend to decrease for higher spins. This is because we normalize the templates at the equilateral limit, while the Legendre polynomials evaluated at $\hat{\mathbf{k}}_1 \cdot \hat{\mathbf{k}}_3 = 1/2$ tend to decrease with the order $s$. The higher spin templates tend to have larger amplitudes, leading to stronger bounds on $\fNL$.

The largest significance level of $1.8\sigma$ is achieved at spin 3, closely tracing that of the orthogonal template. After accounting for the look-elsewhere effect from studying six templates ($s=0,1,2,3,4,6$), the adjusted significance is around $1.2\sigma$.

\paragraph{Massive spin-2 exchange}
\label{sec:results_SO}

\begin{figure}
    \centering
    \includegraphics[width=0.5\textwidth]{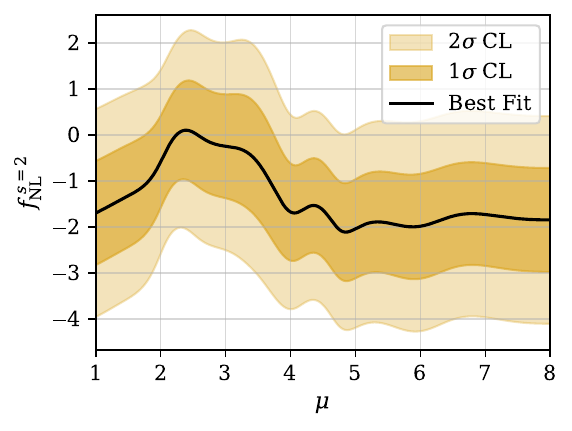}
    \caption{Constraints on the massive spin-2 exchange template \eqref{spin2m} for the mass parameter in range ${1<\mu< 8}$. Each template at $\mu$ has been normalized based on its Fisher information so that the uncertainties are close to $1$. Note the error bars are still estimated from Gaussian simulations and do not exactly equal the optimal value of $1$.}
    \label{fig:SO_constraints_1D_rescaled}
\end{figure}

\begin{figure}
    \centering
    \includegraphics[width=0.46\textwidth]{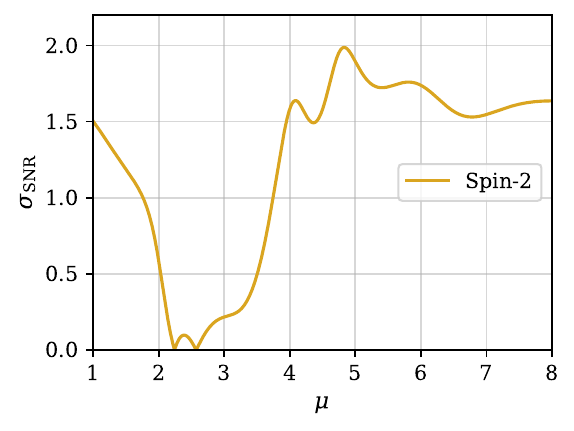}
    \includegraphics[width=0.45\textwidth]{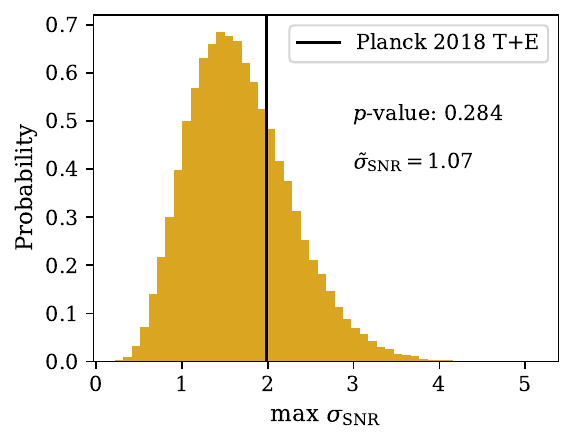}
    \caption{Significance of the best-fit $\fNL$ values for the massive spin-2 exchange template. \textit{Left}: The raw significance $\sigmasnr$ is shown as a function of $\mu$. \textit{Right}: The maximum significance value of 1.99 at $\mu=4.83$ is compared with the theoretical distribution of the maximum statistic. One finds the significance as big as 1.99 roughly $29\%$ of the time. }
    \label{fig:SO_significance}
\end{figure}

The signature of a massive spinning particle with $m \sim H$ is captured in our template \eqref{spin2m}, where we focus on the case with $s=2$. This template has both angular-dependent profiles and squeezed-limit oscillations, which can not be generated in single-field inflation models. We constrain this shape for a range of the mass parameter $\mu$.

We note that the default normalization convention fails for this highly oscillatory template, because the equilateral limit vanishes for some values of $\mu$. While it is possible to normalize using different limits (e.g. set $S(k_*,k_*,2k_*)=1$), we found that the template's overall amplitude varies drastically in most conventions and creates widely varying uncertainties. To avoid unnecessary confusion and to make a fairer comparison of all values of $\mu$ involved, we rescale the template using the Fisher information at each $\mu$. Under this convention, the estimation uncertainty should equal $1$ when the $\fNL$ estimator is optimal.
The rescaled constraints are shown in Figure \ref{fig:SO_constraints_1D_rescaled}. The maximum significance is achieved at $\mu=4.83$ for which the rescaled $\fNL = -2.1 \pm 1.1$, as shown in Figure \ref{fig:SO_significance}. After accounting for the look-elsewhere effect, the adjusted significance is 1.1.

\subsection{Shapes with small sound speeds}
\label{sec:results_EC_LSC_MS}

\paragraph{Equilateral collider shape}
\label{sec:results_EC}

The template in \eqref{eqcol} depends on the mass parameter $\mu$, the sound speed ratio $c_s/c_\sigma$, and the phase parameter $\delta^\mathrm{eq.col.}$. However, as noted in section \ref{sec:cs}, phenomenologically there are only two free parameters $\mu$ and $\phi_{\rm EC}\equiv\mu\log(c_\sigma/c_s)+\delta^\mathrm{eq.col.} $. Furthermore, note that $S^\mathrm{eq.col.}(\phi_{\rm EC}) = \cos\phi_{\rm EC}\ S^\mathrm{eq.col.}(\phi_{\rm EC}=0) + \sin\phi_{\rm EC}\ S^\mathrm{eq.col.}(\phi_{\rm EC} = \pi/2)$ by trigonometric identities. This allows the CMB bispectrum constraints to be placed quickly for all $\phi_{\rm EC}$ with only two evaluations required at  $\phi_{\rm EC}=0$ and $\pi/2$.

\begin{figure}[t]
    \centering
    \includegraphics[width=0.46\textwidth]{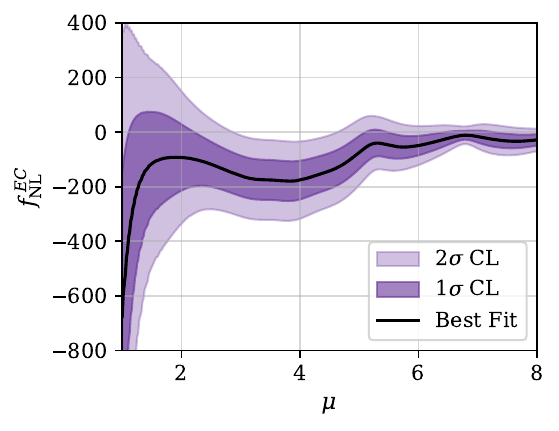}
    ~~\includegraphics[width=0.45\textwidth]{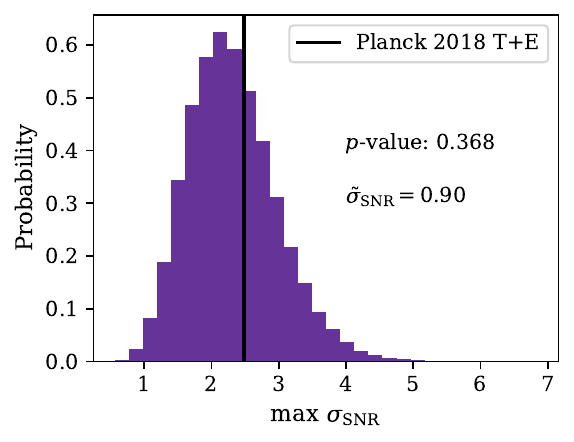}
    \caption{{\it Left}: Constraints on the equilateral collider template \eqref{eqcol} for different mass parameters, when the phase parameter $\phi$ is fixed to the respective best-fit values. The black solid line indicates the best-fit $\fNL$ value at each $\mu$, while the contours represent the 68\% ($1\sigma$) and 95\% ($2\sigma$) confidence levels.
    {\it Right}: The maximum significance of the best-fit $\fNL$ value for the equilateral collider template, compared with the theoretical distribution. One finds the significance as big as 2.5 roughly $37\%$ of the time, so this is fully consistent with random fluctuation. }
    \label{fig:EC_constraints_1D}
\end{figure}

\begin{figure}
    \centering
    \includegraphics[width=0.8\textwidth]{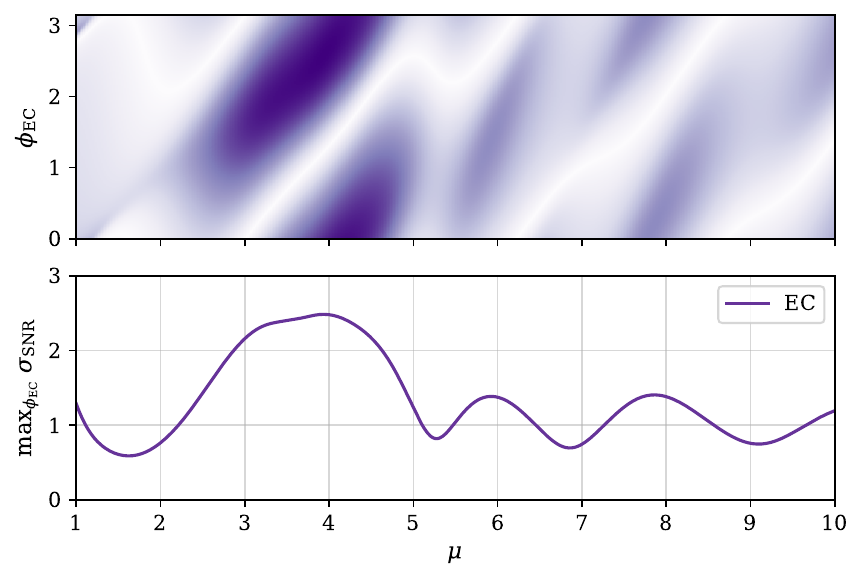}
    \caption{Significance of the best-fit $\fNL$ values for the equilateral collider template \eqref{eqcol}, as a function of the mass parameter $\mu$ and phase $\phi_{\rm EC}=\mu\log(c_\sigma/c_s)+\delta^\mathrm{eq.col.}$. \textit{Top}: The significance $\sigmasnr$ for each $\mu$ and $\phi$ is shown, where darker colours correspond to higher significance. \textit{Bottom}: The significance for which $\phi_{\rm EC}$ fixed to the best-fit value at each $\mu$. Overall, the maximum significance of 2.5 is attained at $\mu=3.94$ and $\phi_{\rm EC}=2.60$. }
    \label{fig:EC_significances_2D}
\end{figure}

The normalization of this template is tricky as the shape function at the equilateral limit becomes $\propto\cos(\mu\log(1/4)+\phi_{\rm EC})$, which can vanish for some parameter choices. We therefore normalize the shape as if the oscillatory part equals $1$, which is equivalent to multiplying \eqref{eqcol} by a factor of $2^{5/2}/3$, or setting $S^\mathrm{eq.col.}(k_*,k_*,k_*)=\cos(\mu\log(1/4)+\phi_{\rm EC})$.

We first show the constraints on $\fNL$ when $\phi$ is fixed to the best-fit values at each $\mu$ in the left panel of Figure \ref{fig:EC_constraints_1D}. Starting with $\fNL=-680 \pm 520$ at $\mu=1$, the maximum significance is reached at $\mu=3.94$ where $\fNL=-178 \pm 72$, eventually reducing to $\fNL=-29 \pm 21$ around $\mu=8$.

The significance is shown as a function of both $\mu$ and $\phi_{\rm EC}$ in Figure \ref{fig:EC_significances_2D}. At a given value of $\mu$, the significance is an oscillatory function of $\phi$ which is always maximised (darkest colour) once in $[0,\pi]$. The maximum significance of $2.5$ is obtained at $\mu=3.94$ and $\phi_{\rm EC}=2.60$, which is a quite significant signal by itself. However, after accounting for the vast parameter space scanned within the 2-parameter search of equilateral collider signals, the adjusted significance reduces to $0.90$, as shown in the right panel of Figure \ref{fig:EC_constraints_1D}.

{The equilateral collider template \eqref{eqcol} becomes highly oscillatory for larger mass parameters $\mu$ while remaining scale-invariant, making it more challenging to accurately expand in terms of separable mode functions \eqref{eqn:basis_expansion}. CMB-BEST utilises a modal basis consisting of Legendre polynomials up to order 29 in each dimension, yielding $~5000$ mode functions up to symmetry. While this basis can accurately represent most templates in this analysis, we found that the expansion accuracy starts to drop after $\mu>5$. The cosine between the template and the modal reconstruction, defined on a tetrapyd, drops to about 0.995 at $\mu=5$. While the basis still captures the main characteristics of the template, the central value $\fNL$ can shift by around $0.1\sigma$ due to imperfect expansion \cite{Planck:2013wtn}. The main difference comes from the increasingly rapid oscillations in the squeezed limit, which is difficult to capture using smooth Legendre polynomials.}

{We therefore like to warn the reader that the exact value of the significances shown in Figure \ref{fig:EC_significances_2D} for $\mu>5$ should be taken with a grain of salt. However, the overall bounds on $f^\mathrm{EC}_\mathrm{NL}$ as shown in Figure \ref{fig:EC_constraints_1D} remain mostly unchanged, and so does our conclusion that $\fNL$ is consistent with zero with the current CMB data.
}

\paragraph{Low-speed collider}
\label{sec:results_LSC}

\begin{figure}
    \centering
    \includegraphics[width=0.6\textwidth]{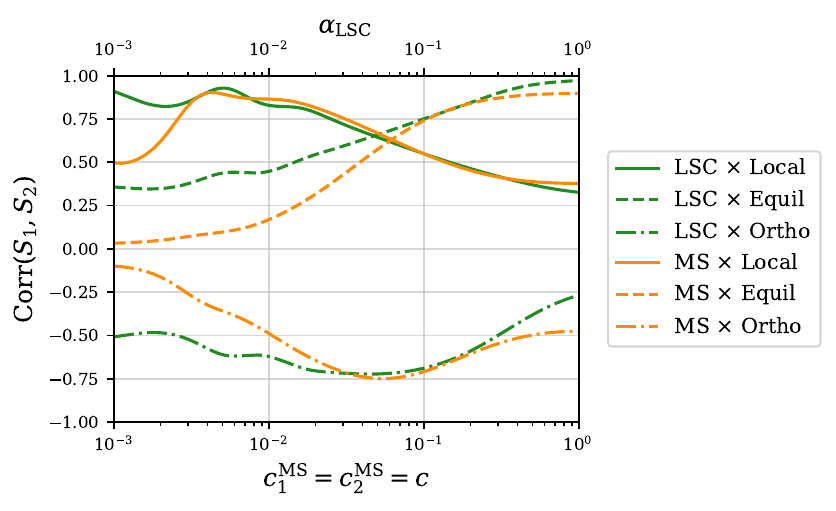}
    \caption{Correlation between the low-speed collider template (LSC, \eqref{lowcs}), multi-speed non-Gaussianity template (MS, \eqref{multics}), and the three standard templates. The correlation is based on the late-time inner product and quantifies the similarity between two templates. As expected from the construction, LSC (MS) templates at $\alpha=10^{-3}$ ($c=10^{-3}$) and $\alpha=1$ ($c=1$) strongly correlate with the local and equilateral templates, respectively. Intermediate values of $\alpha$ ($c$) interpolates between the two. }
    \label{fig:LSC_MS_std_correlations}
\end{figure}

\begin{figure}[b]
    \centering
    \includegraphics[width=0.46\textwidth]{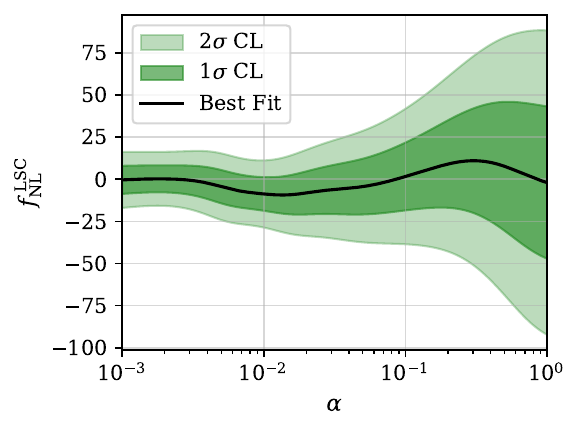}
    \includegraphics[width=0.46\textwidth]{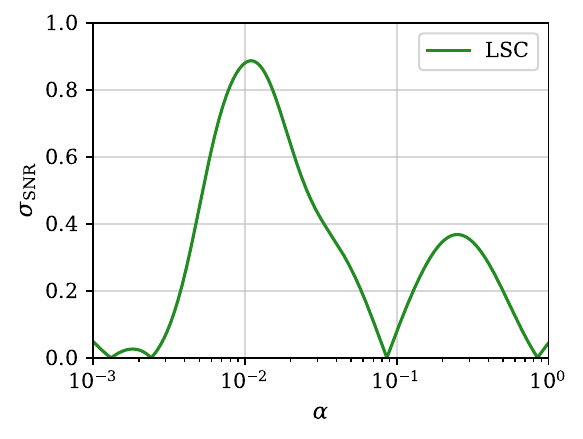}
    \caption{Constraints on the low-speed collider template \eqref{lowcs} for the parameter $\alpha$ varied in $[10^{-3},1]$. \textit{Left}: The best-fit $\fNL$ estimate (black) and the 68\% and 95\% confidence levels (green) are computed independently for each $\alpha$. The amplitude $\fNL$ is consistent with $0$ for all $\alpha$, while being constrained at the $\sigma(\fNL)\sim 10-40$ level. \textit{Right}: The significance $\sigmasnr$ as a function of $\alpha$. The maximum of $0.89$ is obtained at $\alpha=0.011$. }
    \label{fig:LSC_constraints_1D}
\end{figure}

The template \eqref{lowcs} has one free parameter $\alpha\equiv c_s m /H$. For $\alpha=0$ and $1$, it is strongly correlated with the local and equilateral template respectively, as shown in Figure \ref{fig:LSC_MS_std_correlations}. At intermediate values of $\alpha$, the template probes a shape that is in between the two with a moderate correlation with the orthogonal template. Most notable changes in the template appear at small $\alpha$ values, so we present the constraints in log scale in $\alpha$. The template at $\alpha=0$ is not shown here but is very close to $\alpha=10^{-3}$. 

The constraints are shown in Figure \ref{fig:LSC_constraints_1D}.
As we can see, here we have  $\fNL=-0.4 \pm 8.3$ at $\alpha=10^{-3}$ and $\fNL =-2 \pm 45$ at $\alpha=1$. The maximum significance is obtained at $\alpha=0.011$, where  $\fNL=-9 \pm 10$, which remains consistent with $\fNL=0$ within $1\sigma$ level.

\begin{figure}
    \centering
    \includegraphics[width=0.47\textwidth]{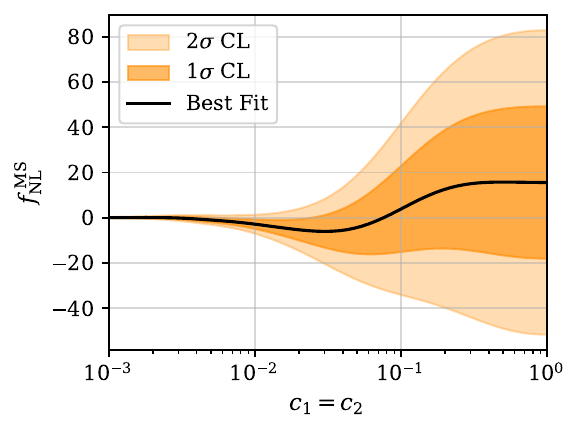}
    \includegraphics[width=0.43\textwidth]{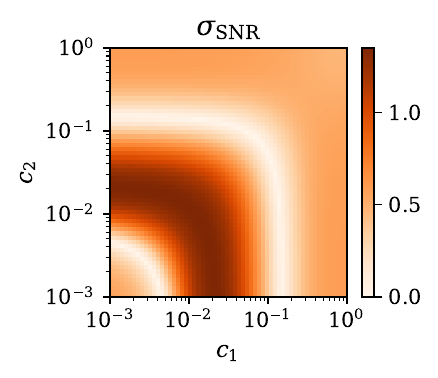}
\\
\includegraphics[width=0.47\textwidth]{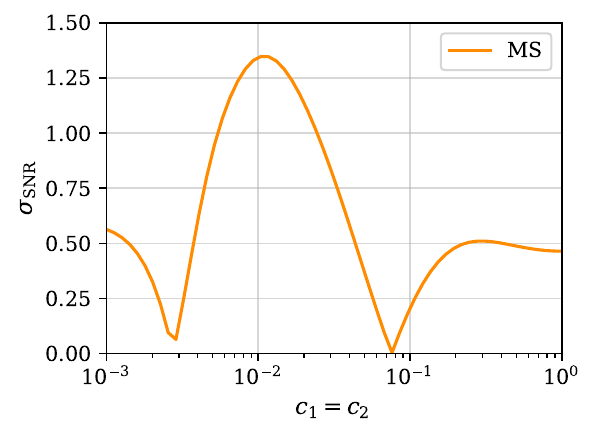}
\includegraphics[width=0.44\textwidth]{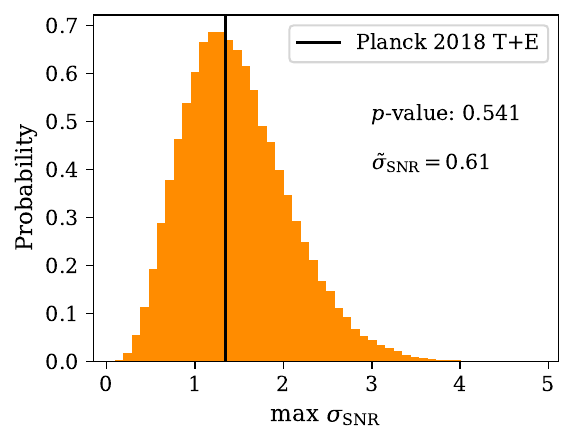}
    \caption{ Constraints on the multi-speed non-Gaussianity template \eqref{multics}. {\it Top Left}: the best-fit $\fnl$ with $1\s$ and $2\s$ contours for $c_1=c_2\in [10^{-3},1]$, while $c_3=1$ is fixed by the overall normalization. {\it Top Right}: The raw significance $\sigmasnr$ is shown as a function of the two sound speeds $c_1$ and $c_2$. 
   {\it Bottom Left}: For $c_1=c_2$, the maximum significance of 1.3 is reached at $(c_1,c_2,c_3)=(0.001,0.021,1)$. 
   {\it Bottom Right}: The maximum significance of 1.3 is compared with its theoretical distribution. One finds the significance as large as 1.3 roughly $54\%$ of the time, so the result is  consistent with statistical fluctuation. }
    \label{fig:MS_significance}
\end{figure}

\paragraph{Multi-speed non-Gaussianities}
\label{sec:results_MS}

The template \eqref{multics} depends on the three sound speed parameters $c_1$, $c_2$ and $c_3$, while the overall normalization fixes one of them. Without loss of generality, here we assume that $c_1,c_2 \le c_3=1$  and normalize it so that $S(k_*,k_*,k_*)=1$. As shown in Figure \ref{fig:LSC_MS_std_correlations}, similar to the low-speed collider shape, the multi-speed template has a big overlap with the local ansatz for $c_1,c_2\rightarrow 0$, and becomes correlated with the equilateral one for $c_1,c_2\simeq 1$.

The top left panel of Figure \ref{fig:MS_significance} shows the constraints for $c_1=c_2 \in [10^{-3},1]$. At $c_1=c_2=10^{-3}$, $\fNL=(0.9\pm1.6) \times 10^{-2}$, while $c_1=c_2=1$ gives $\fNL=16 \pm 34$.  Note that the error bars approach zero in the limit of small sound speed $c_1=c_2 \ll c_3$. This is an artefact of our normalization convention of setting $S=1$ at the equilateral limit. In the squeezed limit, the multi-speed template's amplitude is strongly boosted compared to that of the standard local template, resulting in a larger overall bispectrum and hence tighter $\fNL$ bounds.

The top right panel of Figure \ref{fig:MS_significance} shows the significance where $c_1$ and $c_2$ are allowed to differ. In fact, the maximum significance is attained at $(c_1,c_2,c_3)=(0.001,0.021,1)$ which has $c_1 \neq c_2$, as the colour plot shows. At this point, the constraint is given by $\fNL = -3.1 \pm 2.3$. Note that this significance of $1.3$ is reduced to $0.61$ after adjusting for the look-elsewhere effect, as shown in the bottom right panel.

\section{Summary and Outlook}
\label{sec:concl}

The cosmological collider physics programe provides a unique window on the particle content of the extremely high energy environment of the primordial Universe. It has demonstrated rich phenomenology in PNG, some of which is plausibly detectable in cosmological observations today.
In this paper, we use the legacy release of the Planck CMB observation to constrain the various possible PNG shapes from cosmological colliders. 
Our analysis combines new theoretical understanding from the cosmological bootstrap,
with the recently developed PNG-testing pipeline CMB-BEST. A brief summary of the outcomes is listed below.

\vskip4pt
Based on the bootstrap analysis, we first proposed a set of analytic bispectrum templates for various types of cosmological collider signals. Our investigation incorporates the standard predictions with squeezed-limit oscillations and angular-dependent profiles from massive exchange processes with and without spins. 
Inspired by the full analytical solutions in the bootstrap computation, we propose the approximated scalar seeds and applied weight-shifting operators to derive the simplified templates for scalar bispectra.
These approximate templates successfully reflect the main features of cosmological colliders and significantly enrich the shapes of available PNG.
In addition, we also collect the recently proposed new types of shapes that arise from nontrivial sound speeds. 
These analytical templates (shown in grey boxes in Section \ref{sec:shape}) provide a set of bases for the primordial bispectra that can also be applied in future observational tests for PNG. 

\vskip4pt
Next, we ran the data analysis of these shape templates of cosmological colliders using the CMB-BEST pipeline. We use the temperature and polarization data from the legacy release of \textit{Planck} 2018. 
For each template, we scanned the parameter space to search for best-fit values of $\fnl$ and the corresponding 1$\s$ and $2\s$ constraints. 
We also computed the signal-to-noise ratios to identify the shapes with maximum significance.
The most favoured shape from the \textit{Planck} data is given by the massive scalar exchange II template \eqref{scalarII} with the mass parameter $\mu=2.13$, for which we find  $\fnl=-91\pm40$ at the $68\%$ confidence level.  After accounting the look-elsewhere effect, the largest adjusted significance is $1.8\s$.
For all the shapes, $\fnl=0$ is still consistent with the data.
The main constraints are summarized in Table \ref{tab:summary_constraints}. 
We expect the analysis here to set the stage for
cosmological collider probes in future surveys, which are expected to substantially improve sensitivity and thus have the potential for exciting new discoveries.

\vskip4pt
The hunt for  PNG is a challenging task that needs 
{\it joint} and collaborative efforts between theoretical studies and observational tests.
In the current work, we have endeavoured to overcome the barriers separating the two. Although no evidence of cosmological collider signals is found in the {\it Planck} 2018 data, at this first attempt, our analysis opens up interesting opportunities for future exploration.  Here we list several examples:

\begin{itemize}

\item
In the theory part, the starting point of our bootstrap analysis is the requirement for nearly scale-invariance and weak coupling, which at the leading order leads to the Feynman diagrams in Figure \ref{fig:png} and the corresponding bispectrum shapes of cosmological colliders. However, other possibilities of PNG may appear if we violate these two simple assumptions. For instance, in some models the dominant collider signals arise from loop-level processes; and it is possible to break the scale-invariance during inflation for feature models.
There are many examples in the literature, and they may lead to new types of signatures not captured by our templates here. Meanwhile, as we have shown, CMB-BEST using the general Modal methodology is a convenient tool for testing the non-standard shapes of PNG in the {\it Planck} data. It is encouraging to perform the data analysis for other PNG predictions.
   
    \item 
Second, it will be important to further simplify the bispectrum templates of cosmological colliders. The current choices can represent various typical signatures of massive particles during inflation, however, several of them have rather complicated analytic expressions, and there are still  correlations between these. To search for the primordial signals more efficiently in current and future observations, we are motivated to propose a simple and complete basis of bispectrum templates for various predictions of PNG.

\item
Third, in CMB-BEST, our current choice of basis provides an accurate construction for the primordial templates of cosmological colliders.
An alternative approach is to decompose the CMB bispectrum for the numerical computations.
It would be interesting to continue to cross-validate the CMB-BEST results of this analysis with the Modal estimator used in the main \textit{Planck} analysis. Meanwhile, upcoming CMB experiments such as the Simons Observatory and CMB-S4 will provide pristine measurements of the CMB polarization. Forecasts on the improvements in the cosmological collider constraints from these experiments would be illuminating, especially given the notable (but not yet significant) signals found in our analysis. We leave these for future investigation.

\end{itemize}

\vskip12pt
\paragraph{Note added:} During the preparation of this paper, a relevant work \cite{Cabass:2024wob} appeared on arXiv, which performed the observational test of cosmological colliders using the BOSS data of galaxy surveys.
While we look at the same topic, there are two main differences in the analysis. First, Ref.~\cite{Cabass:2024wob} translated the Planck constraints on the equilateral and orthogonal templates into bounds on collider signals, and mainly focused on the LSS analysis; while  we perform the full CMB
analysis of the PNG templates of collider signals directly using the Planck data.
Second,  Ref.~\cite{Cabass:2024wob} studied the PNG signals from massive scalar exchanges. In addition to that, we scanned the parameter space and also incorporated collider signals from spinning exchanges. For the scalar exchanges, we  have found tighter constraints as shown in  Section \ref{sec:results_QSF_SELM_OC}.

\vskip16pt
\paragraph{Acknowledgements} 

We would like to thank Giovanni Cabass, Xingang Chen, Guilherme Pimentel, Petar Suman for inspiring discussions. DGW is supported by a Rubicon Postdoctoral Fellowship awarded by the Netherlands Organisation for Scientific Research (NWO), and partially by the STFC Consolidated Grants ST/X000664/1 and ST/P000673/1.  Part of this work was undertaken on the Cambridge CSD3 part of the STFC DiRAC HPC Facility (www.dirac.ac.uk) funded by BEIS capital funding via STFC Capital Grants ST/P002307/1 and ST/R002452/1 and STFC Operations Grant ST/R00689X/1.

~

\appendix

\section{Approximated Scalar Seed}
\label{app:seed}

In this appendix, we show the details about the approximated scalar seed proposed in \eqref{Ia} and justify it as a sufficiently accurate approximation of the exact solution $\hat{\mathcal{I}}(u)$.

\begin{figure}
    \centering
    \includegraphics[width=0.49\textwidth]{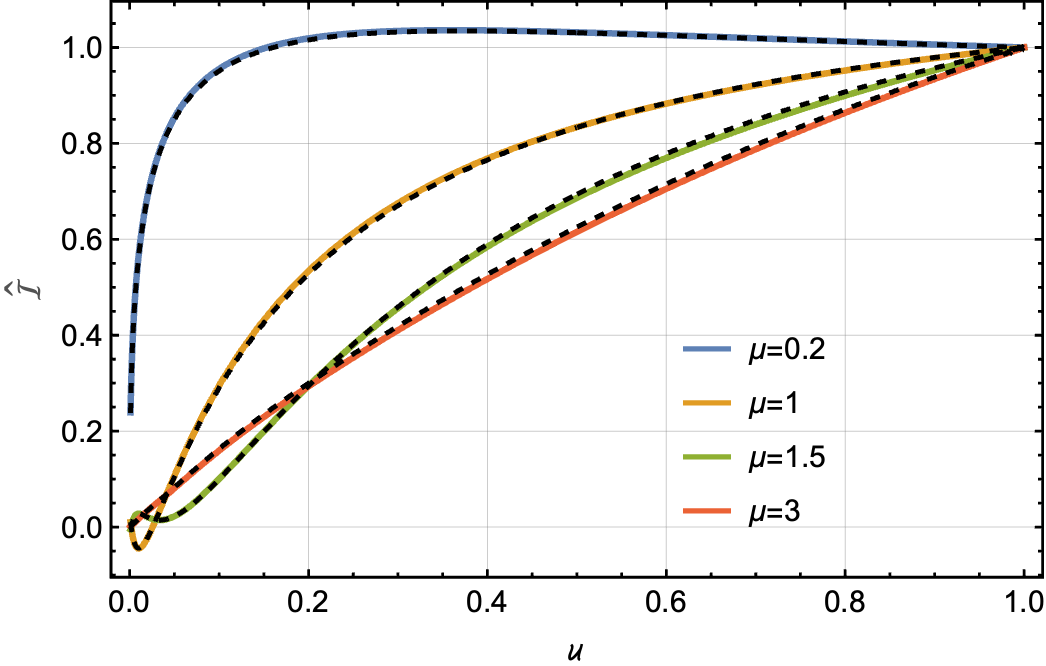}~~
\includegraphics[width=0.495\textwidth]{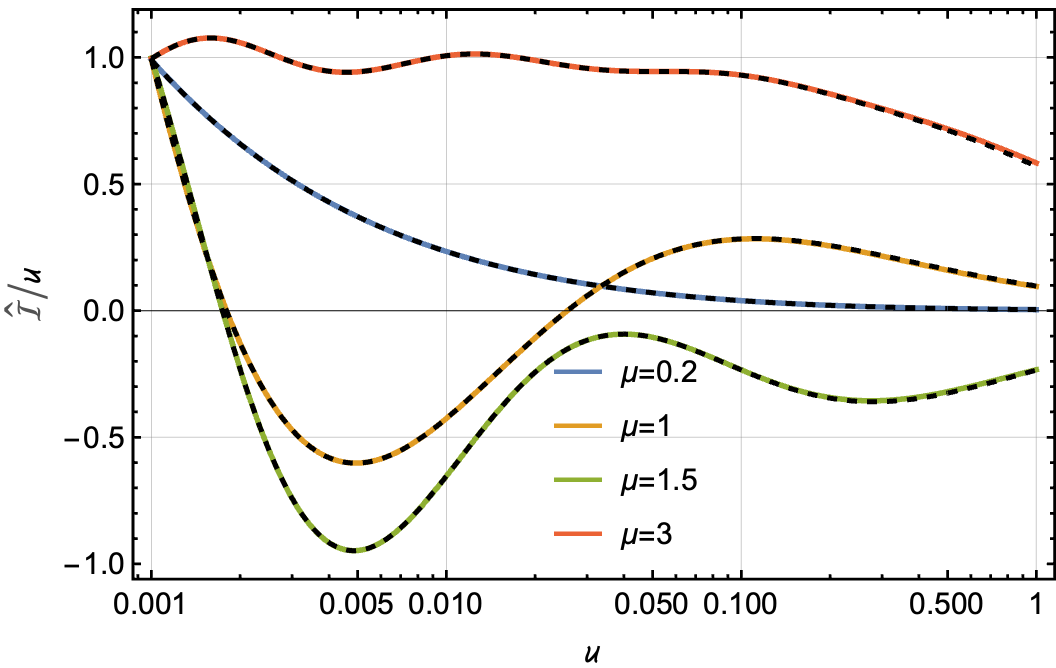}
    \caption{The approximated scalar seed $\hat{\mathcal{I}}$ and $\hat{\mathcal{I}}/u$  with different masses. The dashed lines are the exact solutions from gluing. We adopt different normalizations in these two plots for demonstration.}
    \label{fig:seed}
\end{figure}

Let us first review the exact solution of the boundary differential equation for the three-point scalar seed \eqref{bootstrapeq}. Details of its derivation can be found in Ref. \cite{Pimentel:2022fsc}. The solution has two parts
\be \label{exact}
\mathcal{\hat I} (u) = \mathcal{\hat S} (u)+\mathcal{\hat H} (u)~,
\ee
where the particular piece is given by
\be \label{partisol1}
\mathcal{\hat S} =  \sum_{n=0}^{\infty}c_n u^{n+1}~~{\rm with}~~ c_n = \sum_{m=0}^{\lfloor n/2\rfloor}
\frac{(-1)^n n! / (n-2m)!}{\[\(n+\frac{1}{2}\)^2 + \mu^2\]\[\(n-\frac{3}{2}\)^2 + \mu^2\]...\[\(n+\frac{1}{2}-2m\)^2 + \mu^2\]} ,
\ee
and the homogeneous solution is given by
\be \label{homo-0}
\mathcal{\hat H} = -\frac{1}{2}\sum_{\pm}
B_{\pm} \( \frac{u}{2} \)^{\frac{1}{2}\pm i\mu} {}_{2}F_1 
\[ \frac{1}{4}\pm \frac{i\mu}{2}, \frac{3}{4}\pm \frac{i\mu}{2}; 1 \pm {i\mu} ; u^2 \]~,
\ee
with
\be
B_\pm= \frac{\pi^{3/2}}{\cosh\pi\mu} \(1\mp\frac{i}{\sinh\pi\mu}\)
 \frac{\Gamma(\frac{1}{2}\pm i\mu)}{\Gamma(1\pm i\mu)}~.
\ee
One main advantage of the exact solution is that we can fully understand its singularity structure, which can help us identify where the new physics effects would arise. For instance, there are two nontrivial limits
\be
\lim_{u\rightarrow -1} \mathcal{\hat I}= -\frac{1}{4}\log^2(1+u)~,~~~~\lim_{u\rightarrow 0} \mathcal{\hat I}=-\frac{1}{2}\sum_{\pm}
B_{\pm} \( \frac{u}{2} \)^{\frac{1}{2}\pm i\mu} ~,
\ee
where the first one gives us the total-energy pole, and the second one corresponds to the collider signal as squeezed-limit oscillations.
However, for a practical purpose, this exact solution is less helpful, as it is rather complicated and both two parts contain unphysical singularities at $u\rightarrow 1$
\be
\lim_{u\rightarrow 1} \mathcal{\hat S}= -\frac{\pi}{2\cosh\pi\mu}\log(1-u)~,~~~~~~\lim_{u\rightarrow 1} \mathcal{\hat H}=\frac{\pi}{2\cosh\pi\mu}\log(1-u)~.
\ee
In the final solution these two logarithmic poles would cancel each other and leave the scalar seed function regular in the folded limit $k_3=k_1+k_2$. But in practice, it is difficult to deal with power series with infinite sums and we would also like to get rid of complicated hypergeometric functions in a shape template.
For example, if we want to plot the exact solution  for $0<u\leq 1$, in principle one needs to include all the terms in the power series to obtain a converging result. One way out is to use the exact solution \eqref{exact} in the regime $u\in (0,0.5)$, and then for $u\in (0.5,1)$ we solve the boundary differential equation \eqref{bootstrapeq} around $u=1$ with another converging series expansion, and glue the two solutions at $u=0.5$. These glued solutions are shown by dashed lines in Figure \ref{fig:seed} , which we use as exact results for comparison. 

Next, let us try to find an approximation for the exact solution of $\hat{\mathcal{I}}$, which should be regular at $u=1$ and can match the exact result with good accuracy.
 Our consideration mainly refers to the physical momentum configurations, i.e. $0<u\leq 1$.
We first notice that apart from the singularity at $u=1$, the two parts of the exact solution basically capture two behaviours of the scalar seed: the particular solution provides a $\mu$-dependent equilateral-like shape for the non-squeezed momentum configurations; the homogeneous solution gives us the oscillations in the squeezed limit. This leads our analysis to two different cases.
\begin{itemize}
    \item For $\mu>1$, the homogeneous solution has an overall Boltzmann suppression which leaves the particular solution more dominant for non-squeezed configurations.
    Based on this observation, we propose the approximated scalar seed in \eqref{Ia}, which we  copy paste here
    \be \label{Ia12}
    \hat{\mathcal{I}}_a(u) = \frac{u}{\beta} {(1+u)^{-\frac{\beta}{\beta+2}}}  - \frac{1}{2}\sum_{\pm}B_\pm\(\frac{u}{2}\)^{\frac{1}{2}\pm i\mu}~.
    \ee
    Roughly speaking, the first term there presents an overall shape, while on the top of it the second term gives the small oscillations in the squeezed-limit. In Figure \ref{fig:seed} we compare this template with the exact result for different masses. One can also check that for $m\gg H$, the second term in \eqref{Ia} is exponentially suppressed by the Boltzmann factor $e^{-\pi\mu}$, and the first term leads to $\hat{\mathcal{I}}_a \sim u/(1+u)$ which corresponds to the result of integrating out the $\s$ field in the heavy field limit.\footnote{Note here in the heavy field limit $\mu^2\rightarrow \infty$, only the leading order term gives the expected behaviour in the exact result in \eqref{exact}. 
Beyond the $1/\mu^2$ order, the approximated scalar seed generates $\frac{1}{\mu^{2n}} \log^n(1+u)$, while the correct ones should be $\frac{1}{\mu^{2n}}\frac{1}{(1+u)^{1+n}}$.    
    As we do not attempt to reproduce the total-energy poles in the non-physical momentum configuration $u\rightarrow-1$, these differences in higher-order corrections can be safely neglected.}
    \item For $0 <\mu\leq 1$, the homogeneous and particular solutions become comparable for most kinematic configurations, and one needs precise cancellation between these two to generate the correct result. In this case, the two terms in \eqref{Ia12}  do not agree well with the exact solution. We need to add an extra piece as
    \be \label{fu}
    \hat{f}(u)=-\frac{u}{6\cosh(\pi\mu)} \[ (1-2\mu^4) \log^2(1+u) + \mu^2 u^{1+\frac{8\mu^2}{1+8\mu^2}} \log(1+u) \]~.
    \ee
Figure \ref{fig:seed} also shows the comparison for $\mu=0.2,1$ between $\hat{\mathcal{I}}_a(u)$ and the exact solutions from gluing. The extra term \eqref{fu} is important for these two results to match in the small mass regime $m\simeq 3H/2$.
\end{itemize}

 Furthermore, to derive the bispectrum templates, we need to perform the weight-shifting procedure and add up permutations. Therefore,  it is important to check that the results derived from the approximated scalar seed match  the exact shapes. 
A detailed comparison ensures that the templates shown in Figure \ref{fig:shape-a} and Figure \ref{fig:ortho} are good approximations of the exact results.
Note that, in the $m\simeq 3H/2$ regime, the extra term \eqref{fu} plays a nontrivial role to produce the correct bispectrum shapes. 
Here we also present the extra terms in the scalar exchange templates \eqref{scalarI} and \eqref{scalarIIa}:
\begin{small}
    \bea \label{ddfu}
\tilde{S}_I &\equiv &- k_1 k_2 \partial_{k_1} \partial_{k_2}
\tilde{f} \\
&=& \frac{k_1k_2k_3}{6\cosh(\pi\mu)(k_1+k_2)^3}\[ -{2(2\mu^4-1)}\( \frac{k_3^2}{k_T^2}+ \frac{k_3(4k_T-k_3)}{k_T^2} \log \(\frac{k_T}{k_1+k_2}\)  +\log^2\(\frac{k_T}{k_1+k_2}\) \)  \right. \nn \\
&& \left. + \mu^2 \(\frac{k_3}{k_1+k_2}\)^{\frac{1+16\mu^2}{1+8\mu^2}} \(
\frac{k_3(6k_T-k_3+8\mu^2(8k_T-k_3))}{(1+8\mu^2)k_T^2} + 2\frac{3+68\mu^2+384\mu^4}{(1+8\mu^2)^2}\log \(\frac{k_T}{k_1+k_2}\)
\)\] \nn
\eea
\bea \label{ddfu2}
    \tilde{S}_{II} &\equiv & -\frac{1}{2k_1k_2} (k_3^2 - k_1^2 - k_2^2) {\(1-k_1\partial_{k_1}\)\(1-k_2\partial_{k_2}\)}
\tilde{f} \\
&=&\frac{k_3 \left(k_1^2+k_2^2-k_3^2\right)}{12 \cosh(\pi\mu) k_1 k_2 (k_1+k_2)^4}
\left[ 2\left(2 \mu ^4-1\right) (k_1+k_2) \left(
(k_1^2+k_2^2+3k_1k_2) \log ^2\left(\frac{k_T}{k_1+k_2}\right)\right. \right.\nn\\
&&\left. 
+\frac{k_3 }{k_T^2}\(k_1^3 + k_2^3 +(k_1^2+k_2^2)k_3 + 7k_1k_2(k_1+k_2)+5k_1k_2k_3\) \log \left(\frac{k_T}{k_1+k_2}\right)+\frac{ k_1 k_2 k_3^2 }{k_T^2} \)\nn \\
&& + \mu ^2 k_3 
   \(\frac{k_3}{k_1+k_2}\right)^{\frac{8 \mu ^2}{8 \mu ^2+1}} \left(
   \frac{32 \mu ^2+3}{8 \mu ^2+1} \left(\frac{k_1k_2}{8 \mu ^2+1}  - k_1^2 -k_2^2 -5k_1k_2 
   \right)\log \left(\frac{k_T}{k_1+k_2}\right) \right.\nn \\
&& \left.\left.+\frac{2}{8 \mu ^2+1} \frac{k_1k_2k_3}{k_T}-\frac{k_3}{k_T^2} \Big( k_1^3 + k_2^3 +(k_1^2+k_2^2)k_3 + 11k_1k_2(k_1+k_2)+9k_1k_2k_3\Big) \right)\right]\nn~.
\eea
\end{small}

\vskip6pt
\paragraph{Extension to higher order}
A generalized version of the approximated scalar seed  is needed for deriving the massive spinning exchange templates (see Section \ref{sec:spin} and Equation \eqref{mspins} there). 
These seed functions $\hat{\mathcal{I}}^{(n)}$ satisfy boundary differential equations similar to \eqref{bootstrapeq} but with higher-order source terms \cite{Pimentel:2022fsc}
\be \label{Ineq}
\(\Delta_u + \mu^2+ \frac{1}{4} \)  \hat{\mathcal{I}}^{(n)}(u) = \(\frac{u}{1 +  u}\)^{n+1}~,
\ee
with $n\leq 0$ being integers.
For $n=0$, we return to the scalar seed function in \eqref{I0} that is used for the massive scalar exchange bispectrum.
For $n>0$, this index is associated to the spin of the mediated particle as shown in \eqref{mspins}.
The exact solutions of $\hat{\mathcal{I}}^{(n)}$ and details about their analytic structures can be found in Ref.~\cite{Pimentel:2022fsc}.
Here we propose the following simplified approximation 
\be \label{Ina}
\hat{\mathcal{I}}^{(n)}_a = \frac{u^{n+1}}{\beta} {(1+u)^{-\frac{(n+1)\beta}{\beta+2(n+1)}}}  - \frac{1}{2}\sum_{\pm}B^{(n)}_\pm\(\frac{u}{2}\)^{\frac{1}{2}\pm i\mu} ~,
\ee
with
\begin{small}\be
\beta^{(n)}=\mu^2+\(\frac{1}{2}+n\)^2~, ~~~~~~ B^{(n)}_\pm= \frac{\pi^{1/2}}{2^n}\frac{\Gamma\(\frac{1}{2}+n+i\mu\)\Gamma\(\frac{1}{2}+n-i\mu\) }{\Gamma\(n+1\)} \(1\mp\frac{i}{\sinh\pi\mu}\)
 \frac{\Gamma(\frac{1}{2}\pm i\mu)}{\Gamma(1\pm i\mu)}~.
\ee\end{small}Like \eqref{Ia12}, these approximated scalar seeds provide accurate templates for the exact solutions with $\mu>1$.
In the main text, we substitute \eqref{Ina} into \eqref{mspins} to derive the massive spin-2 exchange template in \eqref{spin2m}.
{ We have also checked that the approximated shape templates shown in Figure \ref{fig:spin2} agree very well with the exact ones from the bootstrap computation.}

\bibliographystyle{utphys}
\bibliography{references.bib}
\end{document}